\documentclass[iop,numberedappendix]{emulateapj}

\newcommand{\nthp}{N$_2$H$^+$}

\usepackage{natbib}
\usepackage{enumerate}
\usepackage{color}
\usepackage{threeparttable}
\usepackage{tablefootnote}
\usepackage{courier}

\usepackage{subfigure}
\usepackage{graphicx}

\begin{document}
\title{Constraining the X-ray and Cosmic Ray Ionization Chemistry of the TW Hya Protoplanetary Disk: Evidence for a Sub-interstellar Cosmic Ray Rate}
\shorttitle{X-ray and Cosmic Ray Ionization of TW Hya}
    \shortauthors{Cleeves et al.}

 \author{L. Ilsedore Cleeves\altaffilmark{1}, Edwin A. Bergin\altaffilmark{1}, Chunhua Qi\altaffilmark{2}, Fred C. Adams\altaffilmark{1,3}, Karin I. {\"O}berg\altaffilmark{2}}

\altaffiltext{1}{Department of Astronomy, University of Michigan, 1085 S. University Ave., Ann Arbor, MI 48109}
\altaffiltext{2}{Harvard-Smithsonian Center for Astrophysics, 60 Garden Street, Cambridge, MA 02138}
\altaffiltext{2}{Department of Physics, University of Michigan, 450 Church St, Ann Arbor, MI 48109} 

\begin{abstract}
 We present an observational and theoretical study of the primary ionizing agents (cosmic rays and X-rays) in the TW Hya protoplanetary disk.  We use a set of resolved and unresolved observations of molecular ions and other molecular species, encompassing eleven lines total, in concert with a grid of disk chemistry models.  The molecular ion constraints comprise new data from the Submillimeter Array on HCO$^+$, acquired at unprecedented spatial resolution, and data from the literature, including ALMA observations of N$_2$H$^+$.  We vary the model incident CR flux and stellar X-ray spectra and find that TW Hya's HCO$^+$ and N$_2$H$^+$ emission are best fit by a moderately hard X-ray spectra, as would be expected  during the ``flaring'' state of the star, and a low CR ionization rate, $\zeta_{\rm CR}\lesssim10^{-19}$ s$^{-1}$.   This low CR rate is the first indication of the presence of CR exclusion by winds and/or magnetic fields in an actively accreting T Tauri disk system. With this new constraint, our best fit ionization structure predicts a low turbulence ``dead-zone'' extending from the inner edge of the disk out to $50-65$~AU.  This region coincides with an observed concentration of millimeter grains, and we propose that the inner region of TW Hya is a dust (and possibly planet) growth factory as predicted by previous theoretical work. 
\end{abstract}


\section{Introduction} 
Gas-rich circumstellar disks around young stars are the formation sites of planetary systems.  The physical conditions of this circumstellar material, including density, temperature, and ionization, all play an important role in setting the dynamical and chemical properties of the disk.   In particular, ionization has a central role in governing disk turbulence and chemistry within the cold ($T<100$~K) planet-forming gas.   The turbulence of disks with masses comparable to our own minimum mass solar nebula \citep[$\lesssim0.05$~M$_\odot$;][]{weidenschilling1977} is thought to be driven by  the magnetorotational instability \citep[MRI; e.g.,][]{velikhov1959,balbus1991,stone1996,wardle1999,sano2000,sano2002,fleming2003,baistone2011}. MRI requires the disk to be sufficiently ionized such that the bulk, predominantly neutral gas can couple to the magnetic field lines, thereby ``stirring'' the gas. Regions of the disk quiescent to such turbulence, i.e., ``dead-zones,'' have been posited as safe-havens for efficient planetesimal formation \citep{gressel2012}, as well as an efficient ``stopping mechanism'' against Type I and II migration \citep[e.g.,][]{matsumura2005,matsumura2007}.  With regards to molecular composition, ionization drives the most efficient chemical processes in the cold, dense regions of disks, both in the gas by ion-neutral chemical pathways \citep{herbst1973} and through ionization-derived hydrogenation reactions on ice-coated grain surfaces \citep{tielens1982,hhl,garrod2008}.  For the same reason, ionization plays a pivotal role in facilitating (or hindering) deuterium fractionation reactions in the gas or on cold grain surfaces \citep{aikawa1999,cleeves2014wat}.  Consequently, ionization is central to the chemical and physical fate of protoplanetary disks and ultimately the planets they form.

The primary sources of dense gas ionization in disks are X-rays, cosmic rays (CRs) and the decay of short-lived radionuclides (SLRs).   Classical T Tauri (CTT) stars are exceptionally X-ray bright \citep[$10^{28}$~erg s$^{-1}$ cm$^{-2} \lesssim L_{\rm XR} \lesssim 10^{34}$~erg s$^{-1}$ cm$^{-2}$;][]{feigelson2002} and often time-variable sources.   X-ray flaring activity in CTTs is commonly associated with an overall hardening of the X-ray spectrum, where relatively more energy is output at  $E_{\rm XR}\gtrsim2$~keV \citep{skinner1997}. These harder X-ray photons are particularly important in setting the disk ionization as they are not easily impeded by intervening material between the star and the disk (i.e., by a stellar/disk wind) and are not as efficiently stopped within the disk itself compared to less-energetic $E_{\rm XR}\sim1$~keV photons  \citep{glassgold1997}.  

 In very dense gas ($n_{\rm H_2}\gtrsim10^{9}$~cm$^{-3}$), where X-rays are strongly attenuated,  the primary sources of ionization available are external galactic CRs and the internal decay of short-lived radionuclides (SLRs).   In the dense interstellar medium, CRs ionize at a rate exceeding a few times $10^{-17}$~s$^{-1}$ and perhaps an order of magnitude or more higher in the diffuse gas \citep{indriolo2012}.  However, in the presence of stellar winds and/or magnetic fields, the incident CR rate can be substantially reduced by orders of magnitude \citep{dolginov1994,cleeves2013a,padovani2013,fatuzzo2014}.

SLRs are provided by massive stars that enrich the dense molecular gas from which young stars (and disks) form.  In addition, certain species including $^{36}$Cl and, to some extent, $^{26}$Al \citep[e.g.,][]{gounelle2001} can be provided by grain-surface spallation via energetic particles from the central star; however, the dominant species at 1~Myr, $^{26}$Al, is primarily formed by external sources \citep[see reviews by][and references therein]{adams2010,dauphas2011}.  The presence of SLRs in the young Solar Nebula is inferred from the Solar System's meteoritic record, but the contribution towards disk ionization from SLR decay is uncertain, owing to unknown initial abundances, inherent time-decay over the lifetime of the disk, and an uncertain ``injection'' point prior to or after disk formation, or perhaps both \citep[e.g.,][]{ouellette2007,ouellette2010,adams2014}. SLR ionization may also be reduced by the escape of ionizing agents, i.e., the decay products, from the tenuous outer disk \citep{cleeves2013b}.

 All of these effects act together to make a rich ionization environment with substantial spatial variation. 
  The individual importance of each of these physical processes has been debated for decades \citep{gammie1996,igea1999}.  For the gas thermal structure in the upper layers, understanding the heating contribution by X-rays, UV irradiation and to a lesser extent the CR flux is necessary \citep[e.g.][]{glassgold2004,jonkheid2004,kamp2004,gorti2009,glassgold2012,bruderer2012}.  In determining the extent of the disk that is unstable to MRI (i.e., is turbulent), the question is often what minimum ionization fraction is required.  CRs, given their ability to penetrate disk gas down to $\lesssim100$~g~cm$^{-2}$, were the original focus of \citet{gammie1996} in determining the thickness of the MRI active layer.  However, given the intensity of the young star as an X-ray source and including the important effects of X-ray scattering, \citet{igea1999} argued that the disk could be turbulent everywhere beyond 5~AU (inside of which X-rays are too highly attenuated) even in the absence of CRs.   With the arrival of spatially and spectrally resolved high signal-to-noise data on molecular ions, many of these questions should be possible to resolve through the coupling of detailed models and observations. 
 
  In \citet{cleeves2014par}, we explored the sensitivity of disk ion chemistry to different ionization scenarios using a generic  T Tauri disk model.  In the present paper, we apply these results to a particular protoplanetary disk, TW Hya.  TW Hya's proximity \citep[$d=55\pm9$~pc;][]{webb1999,vanleeuwen2007}, face-on inclination \citep[$i\sim7\pm1^\circ$;][]{qi2004}, and general isolation from ambient molecular gas \citep{rucinski1983,feigelson1996,hoff1998,tachihara2009}, together provide a clear view into the chemical and physical properties of TW Hya's circumstellar material.  This favorable orientation combined with a rich observed gas-phase chemistry \citep[CN, HCN, DCN, H$_2$O, HD, and H$_2$CO;][]{kastner1997,vanzadelhoff2001,qi2008,hogerheijde2011,oberg2012,bergin2013,qi2013} and numerous detections of molecular ion emission \citep[\nthp\, HCO$^+$, H$^{13}$CO$^+$, and DCO$^+$;][]{kastner1997,vanzadelhoff2001,wilner2003,vandishoeck2003,qi2008,qi2013} make for fertile ground to study the disk's ionization chemistry. In this work, we combine new and archival data with detailed theoretical simulations of disk ionization chemistry to unravel the origin of the molecular ion emission along with its distribution within the TW Hya protoplanetary disk.  
  
  In \S\ref{sec:obs}, we present the observational constraints, including new data from the Submillimeter Array (SMA).  We describe the gas and dust physical model along with the chemical code in \S\ref{sec:model}.  In \S\ref{sec:ionization} we outline the grid of ionization parameters, where we specifically vary the shape of the incident X-ray spectrum and incident CR ionization rate.  We calculate chemical models across the grid and perform simulated observations (\S\ref{sec:lime}) for direct comparison to the data. We present our findings in \S\ref{sec:results}, discuss their implications for the TW Hya disk structure in \S\ref{sec:discussion} and summarize in \S\ref{sec:conclusions}. 

\section{Observations}\label{sec:obs}

\subsection{Submillimeter Array Data}

\begin{figure*}
\begin{centering}
\includegraphics[width=0.68\textwidth]{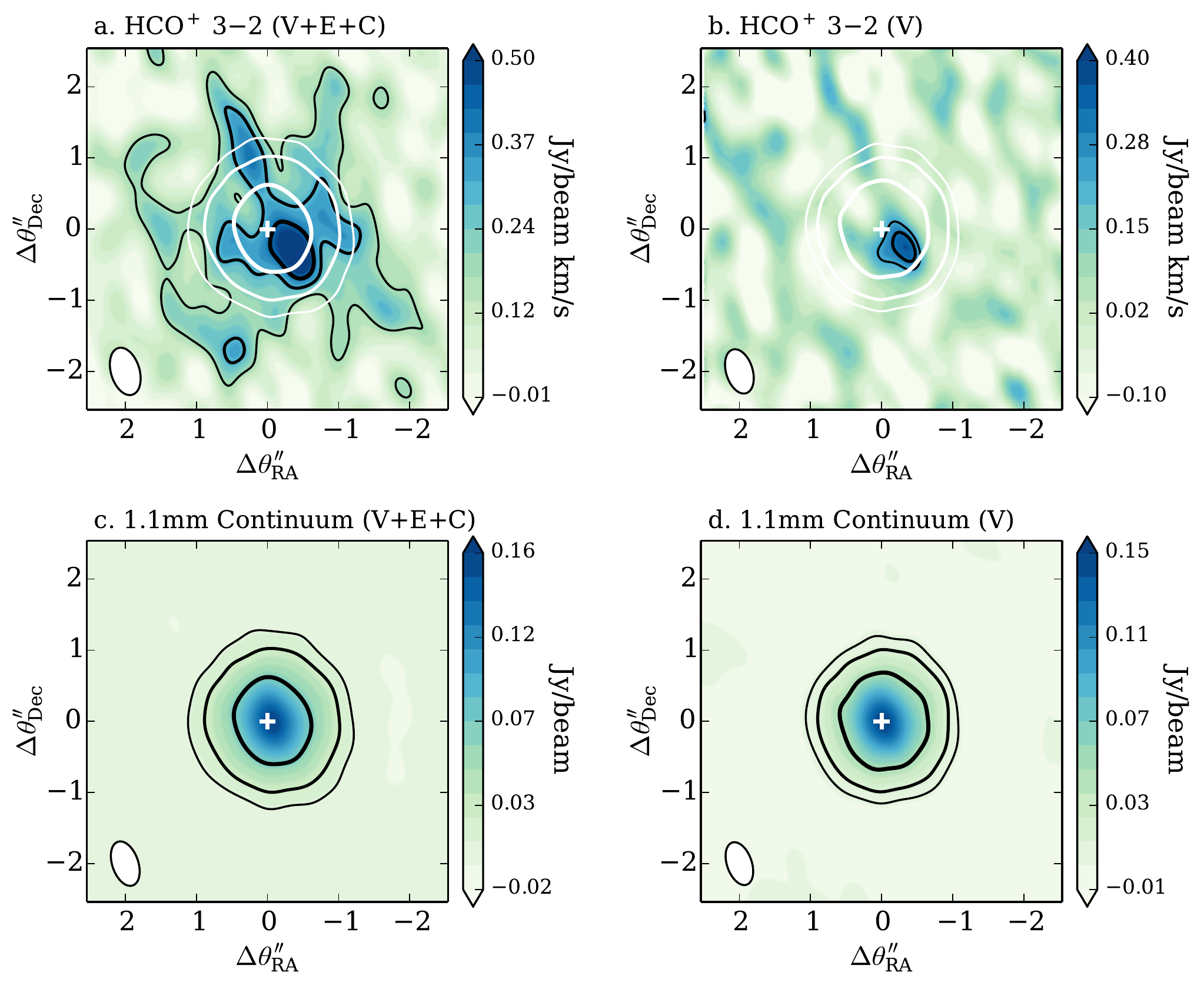} 
\caption{ TW Hya velocity integrated HCO$^+$ ($3-2$) ({\em top}) and 1.1~mm continuum ({\em bottom}) observations. The combined SMA very-extended, extended and compact array data in panels (a) and (c) show the almost symmetric large scale structure of the HCO$^+$ emission, while the very-extended data in panels (b) and (d) show the small-scale deviation from axis-symmetry in HCO$^+$. White crosshairs mark the continuum phase center and overlaid white contours trace the continuum contours. Black contours mark  3$\sigma$, 5$\sigma$ and 7$\sigma$ in the HCO$^+$ panels with 1$\sigma$=60~mJy~beam$^{-1}$~km~s$^{-1}$ ({\em left}) and 85~mJy~beam$^{-1}$~km~s$^{-1}$ ({\em right}). Continuum contours are 3$\sigma$, 10$\sigma$ and 30$\sigma$, where 1$\sigma=2.3$~mJy~beam$^{-1}$ ({\em left}) and 1$\sigma=1.6$~mJy~beam$^{-1}$ ({\em right}).  The beam is shown in the lower-left of each panel. \label{fig:hcopdata}}
\end{centering}
\end{figure*}

The HCO$^+$ (3-2) observations of TW Hydra were made with the SMA \citep{ho2004} located atop Mauna Kea on April 12, 2012 in the very extended (VEX) configuration with 7 antennas under excellent sky conditions, $\tau_{225\rm{GHz}}\lesssim0.04$.  Titan was used for flux calibration, the quasars J1147-382 and J1037-295 were used for gain calibration and 3C279 for passband calibration.  The line was observed with 256 channels in chunk s04, where the chunk width is 104 MHz, corresponding to a velocity resolution of $\Delta v=0.46$~km~s$^{-1}$.  All data were phase- and amplitude-calibrated using the MIR software package\footnote{http://www.cfa.harvard.edu/$\sim$cqi/mircook.html}. All continuum and spectral line maps were then generated and CLEANed using the MIRIAD software package with natural weighting. The synthesized VEX beam for the HCO$^+$ observation is $(\theta_{\rm maj}\times\theta_{\rm min}) = (0.64''\times0.36'')$ with a position angle PA $=17.3^\circ$, where the RMS noise on the line is 85~mJy/beam.   We have also combined the new VEX HCO$^+$ track with shorter spacing data from compact and extended tracks \citep[][observed between 2005-2006]{qi2008} to improve the imaging fidelity.    The synthesized beam for the combined measurement set is $0.69''\times0.39''$, PA $=16.9^\circ$, with an RMS of 60~mJy/beam. The 1$\sigma$ continuum sensitivity at 267 GHz is 1.6~mJy~beam$^{-1}$ ($0.63''\times0.35''$ beam) and 2.3 mJy~beam$^{-1}$ ($0.65''\times0.37''$ beam) beam in the VEX-only and combined observations, respectively.  The decrease in sensitivity in the latter is a result of poorer atmospheric conditions for the extended and compact tracks.  Figure~\ref{fig:hcopdata} shows the velocity integrated line flux for the combined  and VEX-only data sets.

\begin{figure}
\begin{centering}
\includegraphics[width=0.36\textwidth]{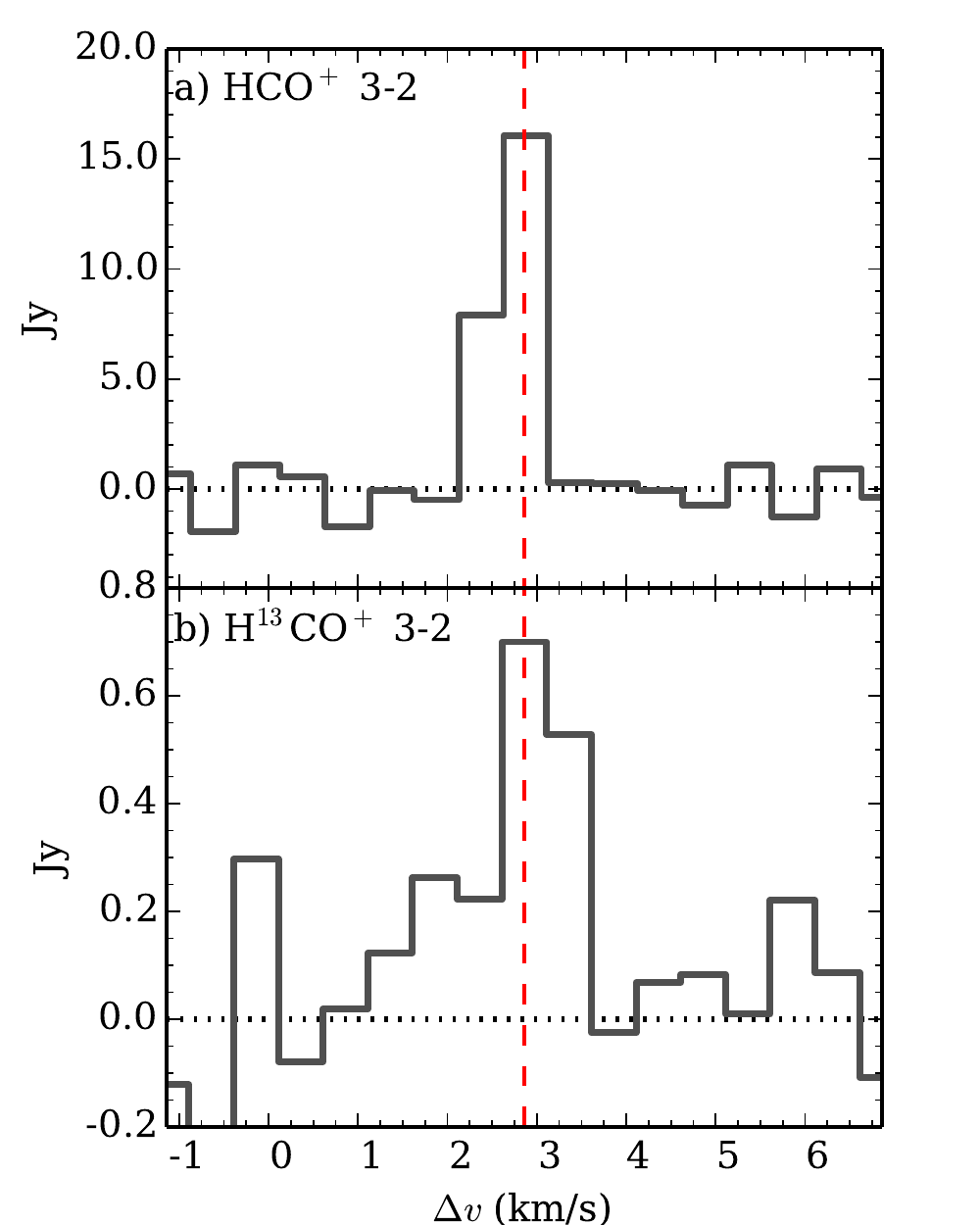} 
\caption{Spatially integrated spectra over an 8$''$ region, where the vertical red line indicates TW Hya's intrinsic velocity $V_{\rm LSR}$=2.86~km~s$^{-1}$. {\em Top:} HCO$^+$ ($3-2$) from the combined (V+E+C) data set. {\em Bottom:} H$^{13}$CO$^+$ ($3-2$) from the extended configuration observations. \label{fig:hcopspec}}
\end{centering}
\end{figure}

Observations of the H$^{13}$CO$^+$ ($3-2$) line were made on 2014 April 8 using six out of eight 6-m antennas of the SMA in the extended configuration with projected baselines ranging from 8 to 165 meters.  The tuning was centered on the  H$^{13}$CO$^+$ ($3-2$) line at 260.255339 GHz in chunk S23. The observing loops used J1037-295 as the gain calibrator. The bandpass response was calibrated using observations of 3C279. Flux calibration was done using observations of Titan and Callisto. The derived flux of J1037-295 at the time of the observations was 0.70~Jy.   The spatially integrated H$^{13}$CO$^+$ ($3-2$) and HCO$^+$ ($3-2$) spectra are shown in Figure~\ref{fig:hcopspec}.

\subsection{Archival Data}\label{sec:additdata}

In addition to the new HCO$^+$ ($3-2$) and H$^{13}$CO$^+$ ($3-2$) data, we have compiled molecular ion emission line observations from archives and the literature (Table~\ref{tab:data}). HCO$^+$ ($1-0$) and ($4-3$) line data was extracted from ALMA Science Verification observations. The HCO$^+$ ($1-0$) line was observed in Band 3 on May 13-14, 2011 with ten 12-m antennas for a total of 3.7~hours.  Titan,  3C279, and J1037-295  were used for flux, gain and phase calibration, respectively.   The HCO$^+$ ($4-3$) line was observed in Band 7 on April 22, 2011 in three scheduling blocks for a total of 4.5~hours.  Nine 12-m antennas were available during the observations; however, one antenna had to be flagged.   The same calibrators were used as in the HCO$^+$ ($1-0$) observations. Further information regarding the observations is provided at the ALMA Science Verification website\footnote{https://almascience.nrao.edu/alma-data/science-verification}.  From the publicly available calibrated data, we extract the flux from an 8$''$ region for both lines. 

H$^{13}$CO$^+$ ($4-3$), N$_2$H$^+$ ($3-2$) and ($4-3$)  have been observed with the SMA and ALMA, respectively \citep{qi2008,qi2013,qioberg2013}. Integrated line fluxes were extracted from the spectral image cubes using an 8$"$ box to be consistent with the new and archival HCO$^+$ data. In addition to these molecular line data, we have made use of the published HD, CO and HCN fluxes listed in  in Table~\ref{tab:data} to calibrate and verify the developed TW Hya disk chemistry model.

All reported noise in Table~\ref{tab:data} combines in quadrature the random noise on the data and an absolute flux uncertainty of 15\%. We note that as a result the detections are individually {\em more significant}, i.e., have higher signal to noise as determined from random noise, than the table implies.  See section \S\ref{sec:lime} for a detailed discussion regarding the ``Recover All?'' column.  

\begin{deluxetable*}{lcccc}
\tablecolumns{5} 
\tablewidth{0pt}
\tablecaption{Spectrally integrated TW Hya line fluxes used to constrain the physical model. Reported uncertainties on the molecular ions include statistical errors and a 15\% systematic uncertainty in the absolute flux scale. \label{tab:data}}
\tabletypesize{\footnotesize}
\tablehead{  
Line                                            &  Integrated Line Flux      & Beam                                        & Reference  & Recover all? \\
& (Jy km/s) & (maj $\times$ min, PA)  & &  }
\startdata
HCO$^+$ ($1-0$)                       & $0.85\pm0.14$   &   ($4.2''\times2.9''$, $72^\circ$)       & ALMA 2011.0.00001.SV.   & Y \\
HCO$^+$ ($3-2$)                        & $12.9\pm2.12$         &   ($0.69''\times0.39''$, $16.9^\circ$)  &This work.                       & Y \\
HCO$^+$ ($4-3$)                       & $23.3\pm3.5$   &  ($1.7''\times1.6''$, $18^\circ$)      & ALMA 2011.0.00001.SV. & N \\
H$^{13}$CO$^+$ ($3-2$)           &   $0.7\pm0.18$                 &    ($1.81''\times0.90''$, $15.6^\circ$)      & This work                        & Y \\
H$^{13}$CO$^+$ ($4-3$)           & $1.1\pm0.41$                  &   ($4.1''\times1.8''$, $3.3^\circ$)                        &  \citet{qi2008} &  N\\
N$_{2}$H$^+$ ($3-2$)               & $2.2\pm0.46$                &  ($3.5''\times2.0''$, $10^\circ$) & \citet{qi2013}  & Y \\
N$_{2}$H$^+$ ($4-3$)               & $4.6\pm0.7$            &   ($0.63''\times0.59''$, $-18^\circ$) & \citet{qioberg2013} & N \\
\hline
HD ($1-0$)                                  & $70.6\pm7.8$               &   ($9.4''\times9.4''$)      & \citet{bergin2013} & Y \\
C$^{18}$O ($2-1$)                      & $0.68\pm0.18$               &   ($2.8''\times1.9''$, $-1.3^\circ$)   & \citet{favre2013,qioberg2013} & Y\\
$^{13}$CO ($2-1$)                      &$2.76\pm0.18$                &   ($2.7''\times1.8''$, $-3^\circ$)  & \citet{favre2013,qioberg2013} & Y\\
HCN ($3-2$)                                &$8.5\pm1.7$                   &    ($1.6''\times1.1''$, $-0.5^\circ$)        &  \citet{qi2008} & Y
\enddata
\end{deluxetable*}

\section{Modeling}\label{sec:model}

To evaluate what constraints the molecular ion observations provide on the ionization agents active in the TW Hya disk requires 1) a physical model of the TW Hya protoplanetary disk, 2) a disk chemistry code, and 3) the application of this code to the physical model under a range of ionizing conditions.  Informed by previous model efforts \citep[e.g.,][]{calvet2002,thi2010a,gorti2011,andrews2012,menu2014} and directly building on \citet{bergin2013}, we have constructed a new physical model of the TW Hya disk, which focuses on the aspects of most importance to the molecular ion chemistry, i.e. the distribution of cold gas and small dust grains.  The details of the modeling process are described in Appendix~\ref{app:mod} as well as the sensitivity of our conclusion on the chosen disk structure, and we review only the main features of the model here. 

\subsection{Physical Structure}\label{sec:struct}

 The dust structure is derived from fitting the spectral energy distribution, where we include settling by defining two dust populations of small (atmosphere) and large (midplane) grains, with the latter concentrated near the
midplane, i.e. it is parameterized as having a smaller scale height. The outer disk radius of our full model is taken to be $R_{\rm out}=200$~AU as determined from the CO gas disk and scattered light images tracing the small dust \citep{andrews2012,krist2000,trilling2001,weinberger2002,roberge2005,debes2013}.  We do not account for the observed radial variation between the two dust populations, i.e., the concentration of large grains inside of $R\sim60$~AU  \citep{andrews2012}, since the explored chemistry depends mainly on the small dust grain population, which dominate the total surface area of grains (see \S\ref{sec:chem}).  
 
\begin{figure*}
\begin{centering}
\includegraphics[width=0.76\textwidth]{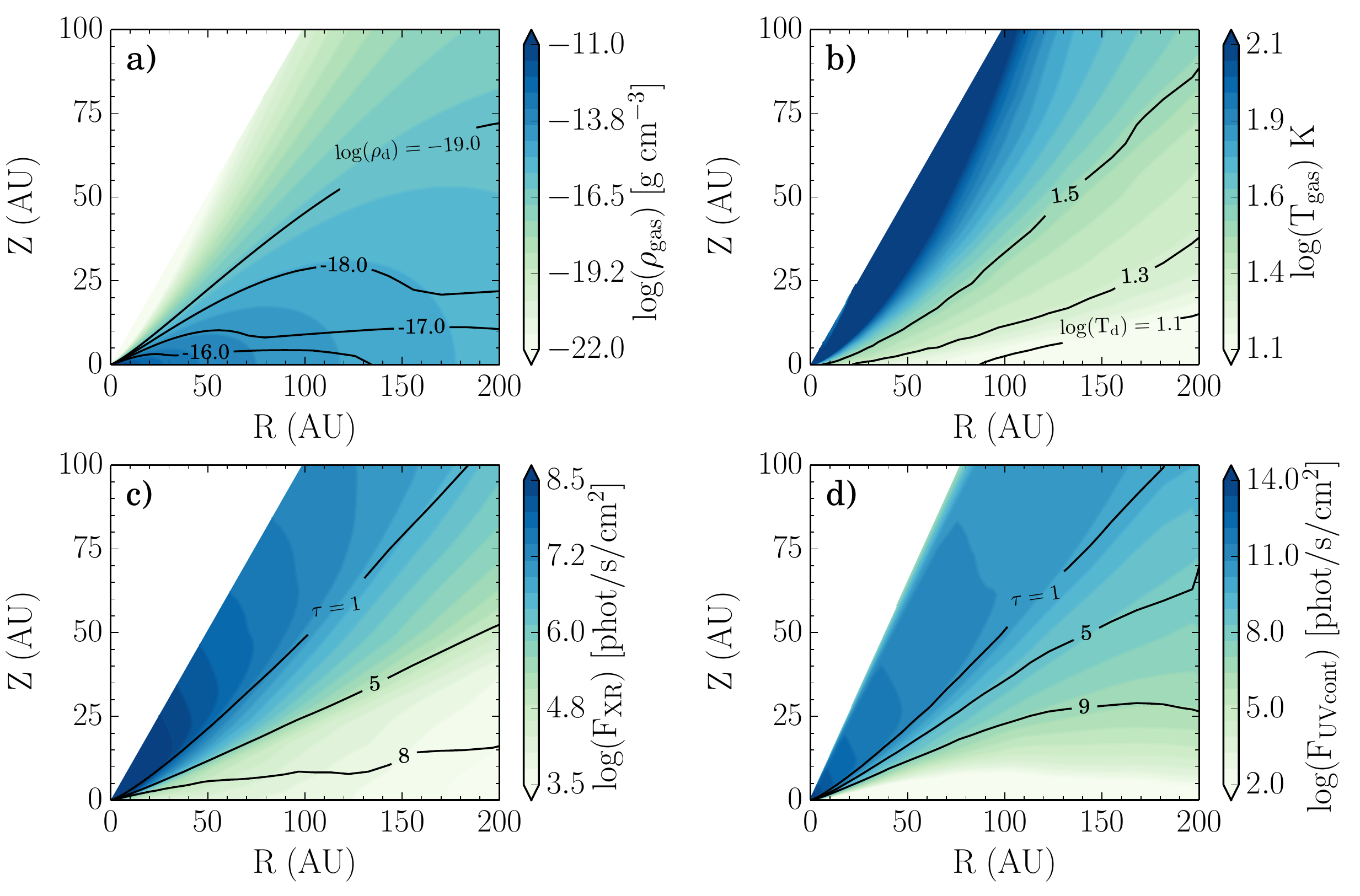}
\caption{TW Hya physical model constrained from the SED and HD observations (see Appendix~\ref{app:addmod}). The filled contours in panels a) and b) are the gas density and temperature, respectively.  The overlaid contour lines in a) and b) show the corresponding dust density and temperature.  Panel c) plots the integrated X-ray flux between 1-20~keV with lines of wavelength integrated optical depth over plotted.  Similarly, panel d) shows the wavelength integrated UV flux along with corresponding UV optical depth.   \label{fig:model}}
\end{centering}
\end{figure*}

We assume the gas is co-distributed with the small grains following a power-law with an exponential taper beyond $R>150$~AU (see Appendix~\ref{app:addmod} for structure dependence) and cut off at an outer disk radius of 200~AU.   We estimate the total mass in gas from the {\it Herschel} detection of HD \citep{bergin2013}, by varying the disk-integrated gas-to-dust mass ratio, i.e., the gas-to-dust ratio calculated from the column density is constant, but the local gas-to-dust ratio varies with height (due to settling). The resulting disk gas mass in our model is $M_g=0.04$~M$_\odot$ with an uncertainty of 0.02~M$_\odot$ (see Appendix~\ref{app:mod}).  The gas temperatures are estimated from the UV radiation field throughout the disk (S. Bruderer, in private communication and Appendix~\ref{app:mod}). The gas and dust distributions of our physical models are shown in Figure~\ref{fig:model}.  
 
\subsection{Disk Chemistry}\label{sec:chem}
We employ a time-dependent chemical code specifically tailored to the disk chemical environment \citep{fogel2011}. The code solves the input reaction network based on the input disk physical parameters and initial chemical abundances.  The baseline chemical reaction network of \citet{fogel2011} is built from the OSU gas-phase network \citep{smith2004} and includes neutral-neutral, ion-neutral, ion recombination with grains/electrons, freeze-out, thermal and non-thermal desorption via UV photons and CRs, photodissociation, photoionization, X-ray induced UV, self-shielding of CO and H$_2$, and water and H$_2$ grain-surface formation.  The expanded network \citep{cleeves2014par} also includes simple deuterium reactions to form H$_2$D$^+$ and HDO, and self-shielding of HD and D$_2$.  The grain-surface formation reactions are also extended in the present work to include additional pathways for grain surface  chemistry in the formalism of \citet{hhl} for a small subset of species (77 reactions total), forming H$_2$/HD/D$_2$, H$_2$O/HDO, H$_2$CO, CH$_3$OH, CH$_4$, CO$_2$, N$_2$, N$_2$H$_2$, HNO, NH$_3$, HCN, OCN, and H$_2$CN. The total network includes 6284 reactions and 665 species.  The initial chemical abundances input into our model are discussed in Appendix~\ref{app:neutral} (see Table~\ref{tab:abun}).  

We constrain the CO and nitrogen abundances using observations of CO and HCN (see Appendix~\ref{app:neutral} for more details).    Regarding CO, motivated by the results of \citet{favre2013} where CO is highly depleted in $\ge20$~K gas by $1-2$ orders of magnitude, we manually change (reduce) the initial CO abundance until our models reproduce the CO isotopologue observations after 1~Myr of chemical evolution.  Chemical processes within the disk naturally remove CO from the gas over time by converting CO into more complex carbon-bearing ices \citep{bergin2014}; however, we find that starting with a ``normal'' CO abundance of $\chi({\rm CO})=10^{-4}$ and allowing the chemistry to evolve over $3-10$~Myr does not alone sufficiently reduce the CO abundance to match the observed C$^{18}$O and $^{13}$CO observations.  This finding implies that some type of additional CO depletion is necessary.  Possible mechanisms include CO chemical conversion to organic ices at even earlier stages, prior to disk formation (i.e., the disk does not start out with $\chi({\rm CO})=10^{-4}$) or, alternatively, freeze-out of CO-derived ices in the disk combined with rapid settling of large, ice-coated grains to the midplane, which can remove carbon and oxygen from the upper layers of the disk.  With our present simple reduction models, we confirm the low gas-phase CO abundance posited in \citet{favre2013}, where we derive a CO abundance of $\chi({\rm CO})=10^{-6}$ relative to H$_2$ (traced by HD).

The chemical calculations are explicitly non-equilibrium, and as such, there is some uncertainty on what ``chemical time'' to adopt when comparing model results with observations.  Time ``zero''  in the chemical code corresponds to a fixed physical structure with uniform input abundances set by Table~\ref{tab:abun} at every location in the disk. As time progresses there is a net gas-phase depletion because there are several ``sinks'' in the network, where energetic He$^+$ atoms break molecular bonds and gradually form more complex species with higher grain-surface binding energies than the parent molecule \citep[for the case of CO, see][and \S\ref{sec:carbon}]{bergin2014}.  

The time for this sequestration process is related to the freeze-out time, which is directly proportional to the typical size of grains assuming a constant gas-to-dust mass ratio \citep[e.g.,][]{aikawa1996}.  Specifically, larger grains have less surface area per unit volume of gas than smaller grains, i.e., the total surface area per unit volume is $\sigma_{g}(r)n_{g}(r_g)\propto r_g^2 r_g^{-3.5}=r_g^{-1.5}$, where $n_g$ is the volume number density of grains and $\sigma_g$ is the surface area of a single grain.  Therefore, as grains grow larger, the surface area of grains drops and molecule collisions with grains are less frequent, slowing down the chemical time for freeze-out.  Likewise, settling removes some fraction of the dust mass from the upper layers and correspondingly, dust surface area, also slowing down the rate of subsequent freeze-out.  Grain growth and settling therefore slow the chemistry's natural tendency to sequester volatiles into more complex ices, and an old disk with large grains can look like a young disk with small grains.  Thus grain growth can reduce the ``chemical age'' as compared to the  physical (stellar) age. 

In the present model, owing to the settled nature of the disk, we reduce the ``chemically active'' surface area of grains to $3\%$ of its standard value, i.e.,  compared to a uniform gas-to-dust mixture (mass ratio of 100) and grains typically $r_g=0.1$~$\mu$m in size, the default assumption of the \citet{fogel2011} model.   Because the typical grain surface area is a single parameter in the model, we cannot increase the surface area in the midplane where the larger ($\ge1$mm) grains will collect.  However, these larger grains contribute negligibly to the surface area compared to the small grains, which are also present throughout the midplane \citep[e.g.,][]{dullemond2004}. This change in grain surface area increases the time-scale for freeze-out by a factor of $\sim30$. 

   Taking these factors into consideration and the unknown age of the TW Hya star and disk system of $\sim3-10$~Myr \citep{barrado2006,vacca2011}, we opt to examine the time-evolved chemical abundances at 1~Myr, when the ion-chemistry has leveled to an approximate steady-state.  Because we cannot take into account the time-evolving grain size and its changing spatial distribution simultaneously with the time evolution of the chemistry along with uncertainties in stellar ages, calibrating the chemical evolution using molecular line observations is a better approach to assessing the current evolutionary state of the TW Hya molecular gas disk. In this light, the present model is verified as appropriate to the current chemical evolutionary stage of TW Hya  using neutral gas constraints, CO and HCN, as described in Appendix~\ref{app:neutral}.

\subsection{Ionization}\label{sec:ionization}
The primary sources of ionization considered in this work are X-rays, CRs, and UV photoionization.  X-rays and CRs are the most important for the ionization of the dense molecular gas, where both are capable of ionizing H$_2$ and helium.  Both the incident CR flux and the implementation of X-rays (motivated by X-ray observations of the source) are the primary free parameters of our modeling efforts and are described in detail below.  We {\em do not} include the ionization contribution from the decay of SLRs, owing to the advanced age of TW Hya \citep[3-12~Myr;][]{webb1999,barrado2006,mentuch2008,vacca2011,weinberger2013} and its relative isolation from recent massive star formation, though see \S\ref{sec:ionresults} for additional discussion.

\subsubsection{CR Ionization}\label{sec:CRs}
The CR ionization rate, $\zeta_{\rm CR}$, is a free parameter in the present study, where we have adopted the results of \citet{cleeves2013a} to construct five input cosmic ray models.  These models consist of empirically motivated CR particle spectra and include self-consistent energy decay with depth (i.e., surface density).  The CR attenuation is taken in the vertical direction for ease of calculation in the chemical models.  In Table~\ref{tab:rates} we provide the incident CR ionization rate at the surface of the disk (prior to attenuation by the gas itself).  
\begin{deluxetable}{lcrrr}
\tablecolumns{2} 
\tablewidth{0pt}
\tablecaption{Incident CR model ionization rates, $\zeta_{\rm CR}$. \label{tab:rates}}
\tabletypesize{\footnotesize}
\tablehead{ Model &  ID &  $\zeta_{\rm CR}$  }
\startdata
 \citet{moskalenko2002}  & M02 & $ 6.8\times 10^{-16}$ s$^{-1}$\\
\citet{webber1998} & W98 & $ 2\times 10^{-17}$ s$^{-1}$\\
Solar System Minimum &SSM & 1.1 $\times 10^{-18}$ s$^{-1}$\\
Solar System Maximum  &SSX & $ 1.6 \times 10^{-19}$ s$^{-1}$\\
T Tauri Minimum & TTM & 7.0 $\times 10^{-21}$ s$^{-1}$
\enddata
\end{deluxetable}
Details regarding the calculation of $\zeta_{\rm CR}$ and the functional form of the decay with surface density can be found in \citet{cleeves2013a}. In summary, the \citet{moskalenko2002} (M02) ionization rate is similar to that determined for the diffuse ISM, the \citet{webber1998} (W98) rate is consistent with values for the dense, molecular ISM, and Solar System Minimum (SSM) and Solar System Maximum (SSX) are the present day CR fluxes at 1~AU.  T Tauri Minimum is an ``extreme'' modulation case, extrapolated from the solar values \citep[see][for details]{cleeves2013a}.

\subsubsection{X-ray Ionization}\label{sec:xrays}
The efficiency of X-rays to ionize a disk depends both on the total flux and the hardness of the spectra. TW Hya has a substantial measured X-ray luminosity of  $L_{\rm XR}\sim2\times10^{30}$~erg~cm$^{-2}$~s$^{-1}$ \citep[e.g.,][]{kastner1999,raassen2009,brickhouse2010}, assuming a distance of $d=55$~pc. This flux has been observed to double or triple  during X-ray flares over periods of hours \citep{kastner2002,brickhouse2010} at a cadence of less than a day \citep{kastner1999}.  

The quiescent state X-ray spectrum is well-fit by a two temperature plasma model with characteristic temperatures of $2-3$ MK and $10-12$ MK \citep{kastner1999,brickhouse2010}, which are associated with X-ray emission from the accretion shock \citep{calvet1998,kastner1999,kastner2002,stelzer2004,brickhouse2010,dupree2012}, and coronal emission from hot plasma  \citep[e.g.,][]{lamzin1999,gudel2007,brickhouse2010}, respectively. During flares, hints of a hard X-ray excess have been observed by \citet{kastner1999}. \citet{kastner2002} and \citet{brickhouse2010} also found that during independent flaring events, harder energy bands and high temperature diagnostics were affected by the flare while the soft component was not.

Despite the exquisite data available on X-ray fluxes and spectra toward TW Hya, its impact on ionization and chemistry is highly uncertain. First, the shape of the spectrum beyond $E_{\rm XR}\sim3$~keV (the very photons that ionize the bulk gas) in both the quiescent and flaring states is not directly known. Second, the variability of the X-ray flux and possibly spectrum occurs on small enough time scales that the chemistry may not be able to reset between flares, and may thus reflect the flared state of TW Hya dependent on the relevant timescales. For ion chemistry the relevant time scale is the electron recombination rate. The typical electron recombination rate coefficient is of order $\alpha_{\rm rec}\sim1\times10^{-7}$~cm$^{3}$~s$^{-1}$.  The electron density at the inner disk surface, $R=30$~AU, $Z=12$~AU (i.e., above the $\tau_{\rm XR}\sim1$; Figure~\ref{fig:model}), is approximately $n_e\sim10$~cm$^{-3}$, resulting in a recombination time of $t_{\rm rec}\sim\left(\alpha_{\rm rec}n_e \right)^{-1}=12$~days.  Therefore, if the flaring time scale is on the order of days, the chemistry may not have time to ``reset'' between flares.  
 
\begin{figure} 
\begin{centering}
\includegraphics[width=0.39\textwidth]{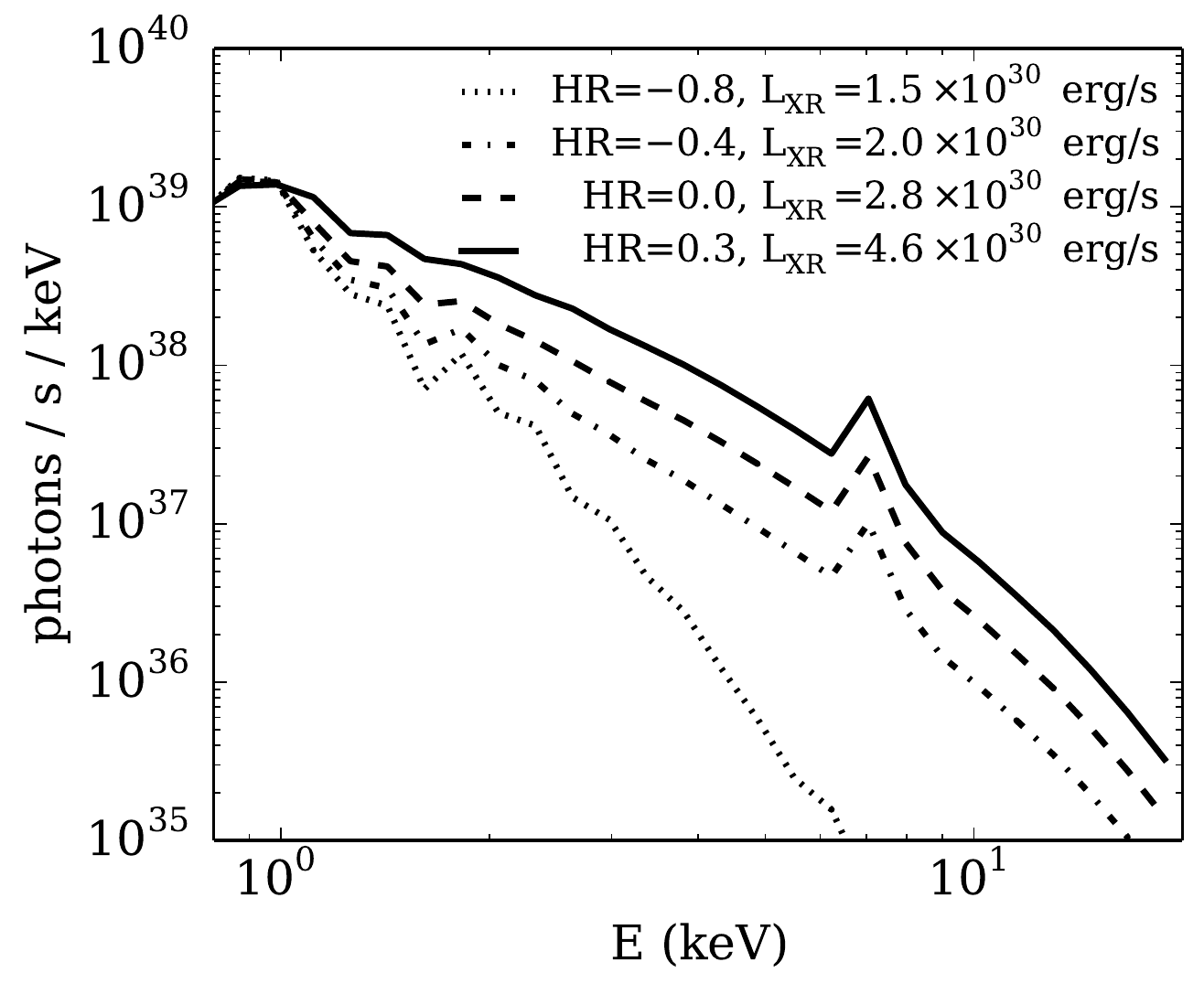}
\caption{Template X-ray spectra for the ionization model grid, where the softest X-ray spectrum (i.e., the quiescent template), $\rm HR=-0.8$, is  motivated by the observations of \citet{kastner1999,kastner2002}, while the other spectra are artificially hardened to simulate different degrees of ``flaring'' spectral states. 
\label{fig:X-rays}}
\end{centering}
\end{figure}

To address this issue, we treat the X-ray flux and spectrum as a free parameter in our model. We construct four low-resolution X-ray templates with XSPEC\footnote{http://heasarc.gsfc.nasa.gov/xanadu/xspec/.}, where the baseline ``quiescent'' X-ray model is the two component plasma model (raymond) derived by \citet{kastner1999}, with 1.7 MK and 9.7 MK components (these components are similar to those found in the detailed spectroscopic study of \citet{brickhouse2010}). The luminosity in the quiescent template is $L_{\rm XR}=1.5\times10^{30}$~erg~cm$^{-2}$~s$^{-1}$.  To approximate the X-ray flaring state, on top of the quiescent spectrum we add an artificial hard component at 4~keV, changing the relative amounts of hard-to-soft X-rays and thus decreasing the overall spectral slope (Figure~\ref{fig:X-rays}).  

We fix the soft X-ray flux at 1 keV across the four spectra based upon the results that the low temperature, softer X-ray component is unaffected by the observed flares.  Between the four spectra considered here, the integrated X-ray luminosity changes by a factor of just $\sim3$ (see Figure~\ref{fig:X-rays} legend); however, the specific flux at the hard X-ray tail ($E_{\rm XR}\gtrsim5$~keV) changes by an order of magnitude. All but the highest energy flaring spectrum are consistent within the error bars of the data presented in \citet{kastner1999}.
We identify the four models by their X-ray hardness ratio, which is defined as the ratio $\left(L_{\rm soft}-L_{\rm hard}\right)/\left(L_{\rm soft}+L_{\rm hard}\right)$ where $L_{\rm soft}$ is the X-ray luminosity between $0.5-2.0$~keV in erg~s$^{-1}$ and $L_{\rm hard}$ is the equivalent between $2.0-10$~keV.  More negative numbers are soft X-ray dominated, while more positive numbers are hard X-ray dominated. 

 We calculate the X-ray propagation throughout the disk for energies of $E_{\rm XR}=1-20$~keV in 1~keV intervals as described in \citep{cleeves2013a} with the Monte Carlo code presented in \citet{bethell2011b}.  The absorption cross-sections are provided in \citet{bethell2011a} and the X-ray scattering is dominated by Thompson scattering.    Next, we apply the template spectra in Figure~\ref{fig:X-rays} to determine the X-ray radiation field as a function of energy and position within the disk, which is the input for the chemical calculations. The X-rays primarily ionize H$_2$ and He, where we adopt the ionization cross-sections provided in \citet{yan1998}, integrated over the local X-ray spectrum.  The same \citet{bethell2011b} code is also used for the UV  calculations \citep[see][for our treatment of the Lyman-$\alpha$ transfer and \S\ref{sec:gas}]{cleeves2014par}. The energy-integrated X-ray and UV field ($930-2000$~\AA) are shown in Figure~\ref{fig:model}c and \ref{fig:model}d, respectively, along with their total optical depth.  
 
 \begin{figure}
\begin{centering}
\includegraphics[width=0.43\textwidth]{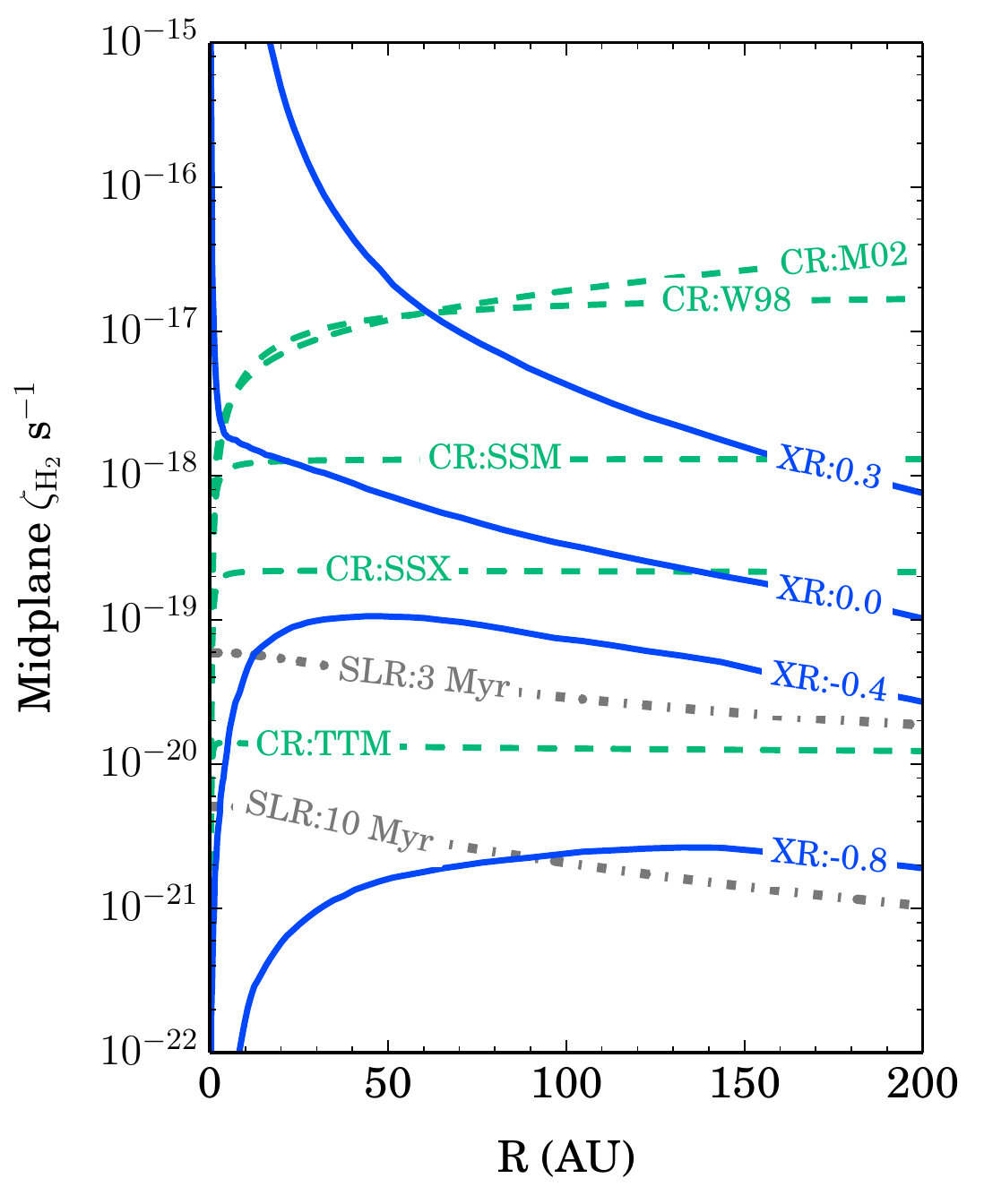}
\caption{H$_2$ ionization rate in the midplane ($z=0$~AU) due to X-rays (XR), cosmic rays (CR) and short-lived radionuclides (SLR). Incident (unattenuated) CR fluxes are listed in Table~\ref{tab:rates}.  X-rays are labeled by their hardness ratio (see also Figure~\ref{fig:X-rays} for the incident spectra). SLR rates are determined from the initial solar nebular abundances with the indicated decay time (and no late stage injection).   \label{fig:ionmods}}
\end{centering}
\end{figure}

\subsection{Line Radiative Transfer and Synthetic Observations}\label{sec:lime}

The set of five CR models (\S\ref{sec:CRs}) and four X-ray models (\S\ref{sec:xrays}) forms a grid of twenty different ionization profiles for a fixed physical (density and temperature) structure (Appendix~\ref{app:mod}).  For each model, we calculate the time-dependent chemistry as described in \S\ref{sec:chem}.  To enable model-data comparison, we then calculate the emergent line emission from the abundances and model physical conditions assuming a distance of $d=55$~pc and, for cases where spatial filtering cannot be excluded, we also simulate the particular observations. 

The line radiative transfer is computed using the LIME code \citep{brinch2010} in non-LTE where available\footnote{Collisional rates for HCO$^+$, H$^{13}$CO$^+$ and N$_2$H$^+$ are compiled at the Leiden LAMDA database \citep{schoier2005}, where the primary literature for the spectroscopic data is \citet{anderson1977,sastry1981,botschwina1993,flower1999}.}. We consider two components to the gas velocity: i) Keplerian rotation around the star based on the stellar mass and viewing geometry ($i\sim7^\circ$) and ii) the gas turbulent velocity, i.e., the doppler-B parameter.  We adopt a turbulent velocity of 40~m~s$^{-1}$ based on the observed upper limit from \citet{hughes2011}. 
The end product of the LIME simulations are data cubes, i.e., two dimensions on the plane of the sky and a third in velocity.   Since carbon and oxygen isotopologues are not considered independently in the chemical network, their abundances are calculated based on 
the main isotopologues using a fixed ratio of $\rm ^{12}C/^{13}C=70\pm10$ and $\rm ^{18}O/^{16}O=540\pm30$ \citep{henkel1994,prantzos1996} appropriate for the local ISM.   For all lines, we simulate the line and continuum emission simultaneously with the dust and gas co-spatial within the framework of the physical model.   We then use self-consistent dust opacities from the radiative transfer and UV modeling, and then subtract off the continuum emission when comparing the line fluxes. 

The calculated line emission fluxes are compared to observations to determine the goodness of fit of the chemical simulations. This can be done directly using the LIME output for lines observed for the observations that have sufficient short spacings to not have any flux resolved out, marked as `Recover all?'='yes' in Table~\ref{tab:data}. Exceptions to this are the $(4-3)$ rotational lines of HCO$^+$, H$^{13}$CO$^+$, and N$_2$H$^+$. For these lines, we make synthetic observations using the \texttt{simobserve} task in CASA\footnote{http://casa.nrao.edu/} and the specific array configurations used for these SMA and ALMA observations. The HCO$^{+}$ ($4-3$) and N$_2$H$^{+}$ ($4-3$) were both observed as part of ALMA Early Science operations, where the former was a Science Verification target and the latter was reported in \citet{qioberg2013}.  Because the array configurations during these observations were non-standard, we used the original data to create the observation specific array configurations for the \texttt{simobserve} task.  From the simulated visibilities we reconstruct the on-sky image using \texttt{clean} with natural weighting.  The simulated beam from the reconstructed images is ($4.1'' \times1.97''$) and ($0.66''\times0.56''$) for HCO$^+$ ($4-3$) and N$_2$H$^+$ ($4-3$), respectively, which are in good agreement with the beams from the observations (see Table~\ref{tab:data}).   Likewise, the CASA simulations for the SMA H$^{13}$CO$^+$ ($4-3$) with a  beam of ($4.1''\times1.97''$) reasonably agrees with the observed beam.

For all lines except HCO$^+$ ($3-2$) and N$_2$H$^+$ ($4-3$), we only compare spatially/spectrally integrated line fluxes.  To determine quality of fit, we combine the uncertainties on the original observations in quadrature with a 10\% uncertainty from the stochastic point-casting uncertainty of the LIME code \citep{brinch2010}. For the uncertainty of the two H$^{13}$CO$^{+}$ rotational lines, we also include a 14.3\% uncertainty for the isotopologue ratio of $^{12}$C/$^{13}$C$=70\pm10$.

For the best spatially resolved data, HCO$^+$ ($3-2$) and N$_2$H$^+$ ($4-3$), we directly compare the observed emission profile on the plane of the sky to the model emission profile.  To assess the fit to the model emission profiles, we measure the difference between the model and observed line fluxes in units of the integrated $\sigma$ (Jy/beam~km~s$^{-1}$) over the angular extent of the disk.  From this difference, we determine the radial difference in units of the uncertainty on the line flux, $\sigma(R)$.   From this profile, we determine a disk-averaged $\sigma$ between the model and observation by integrating over the disk area, $\int2\pi \Delta\sigma(\theta) \theta d\theta$/$\int2\pi \theta d\theta$, which implicitly weights towards the outer disk because it covers more of the total emission area.  For both HCO$^+$ ($3-2$) and N$_2$H$^+$ ($4-3$) we integrate out to $\Delta\theta=3''$ from the stellar position, beyond which the emission drops off substantially.

\section{Results}\label{sec:results}
\subsection{HCO$^+$ Spatial Distribution}
The HCO$^+$ ($3-2$) velocity integrated line emission for the combined data set and the VEX data set are shown in Figure~\ref{fig:hcopdata} together with the simultaneously observed continuum data. The VEX data clearly deviates from axisymmetry, and while less pronounced, this deviation is also recovered in the combined image. The VEX emission excess is separated from the phase center by the size of the $\sim$25~AU beam and seems to be unresolved. In the combined image, the feature shows up  as a $3\sigma$ enhancement as compared to the emission profile across the rest of the disk at the same radial distance.  The emission contributes approximately $\sim15-20\%$ of the overall disk-integrated HCO$^+$ ($3-2$) flux, which is a substantial amount of the total emission confined to a very small region. The possible origin of this asymmetry is discussed in \S\ref{sec:asymmetry}.  While the asymmetry is not the focus of the present paper, equipped with this additional spatial information, we are able to exclude the southwest quadrant from the ionization fitting as it does not reflect the majority of the disk properties.  The three other quadrants are assumed to be representative of the axisymmetric HCO$^+$ ($3-2$) emission profile.  The remaining HCO$^+$ lines, ($1-0$), ($4-3$) and the isotopologues, were observed at lower spatial resolution (see Table~\ref{tab:data}, and as a result, we are unable to correct their emission for such an asymmetry if it is a long-lived feature.  These observations highlight the utility of high spatial resolution observations even for the study of bulk gas chemical properties.  

\subsection{Model Grid Results}\label{sec:modresults}
The contribution of the different X-ray and cosmic ray models to the disk midplane ionization level is shown in Figure~\ref{fig:ionmods}.  We also plot the contribution from two SLR realizations assuming an age of 3~Myr and 10~Myr with Solar System-like initial SLR abundances \citep[see][for details]{cleeves2013b}.  The relative importance of CR and X-rays for H$_2$ ionization clearly depends both on the CR attenuation and the X-ray hardness. In its quiescent state, even the most attenuated cosmic rays dominate the X-ray ionization level throughout the disk. Moderate hardening/flaring of the X-ray spectra results  in an X-ray dominated ionization profile in the inner disk, and for the hardest X-ray spectra, X-rays drive ionization throughout the disk for all cases with any CR attenuation.  SLRs only contribute to the ionization level if the CRs are extremely attenuated and the star displays no X-ray flaring activity, justifying their exclusion from the model calculations. 

\begin{figure}
\begin{centering}
\includegraphics[width=0.483\textwidth]{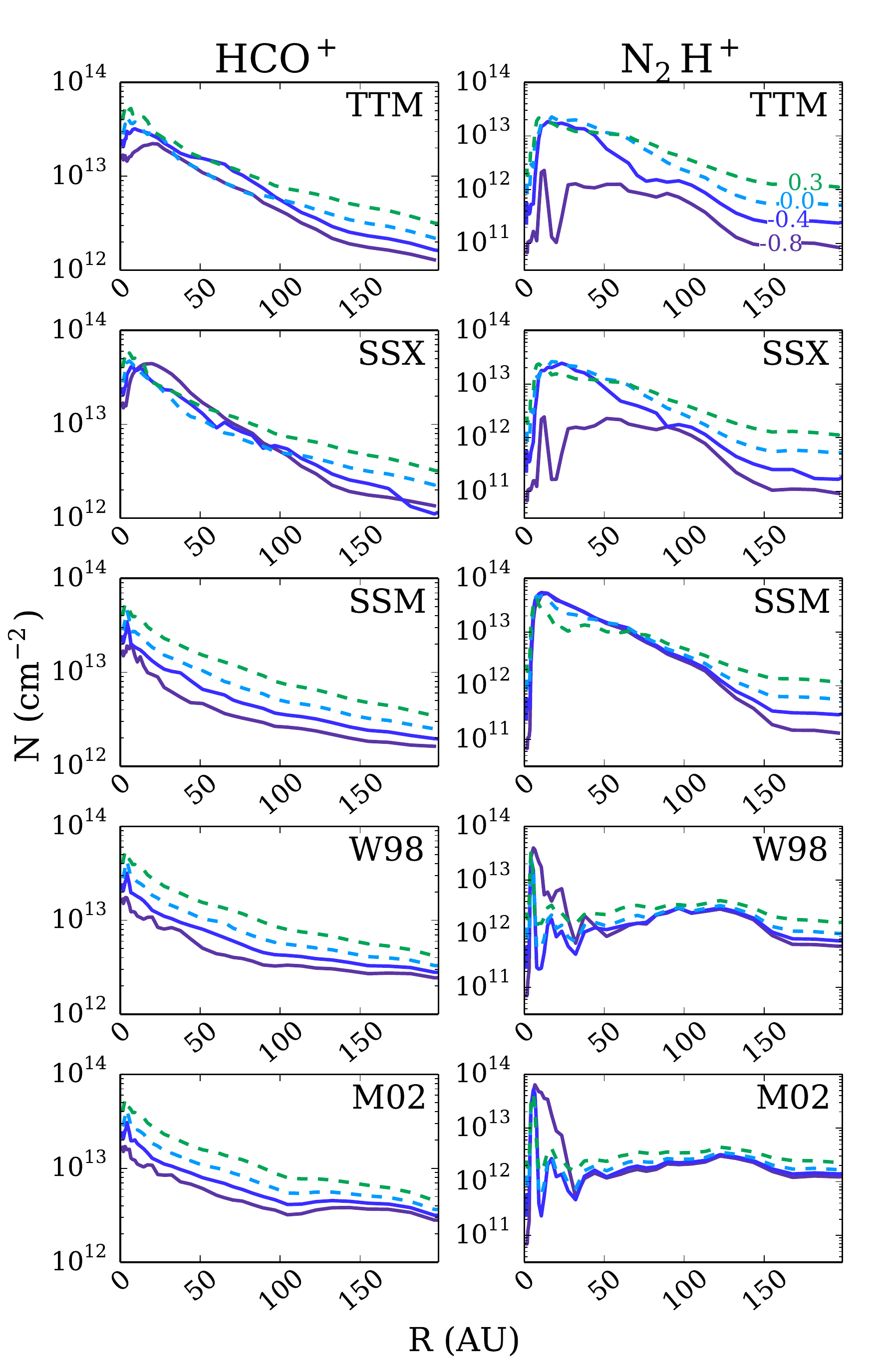}
\caption{Vertically integrated column densities of HCO$^+$ (left) and N$_2$H$^+$ (right) versus disk radius.   Individual panels show increasing CR ionization rates from top to bottom. In each panel,  variations due to changes in the X-ray spectral hardness are shown by the different lines as labeled in the top right panel.  The dashed lines indicate hard X-ray spectra, 0.0 and 0.3, while the solid lines are the softer X-ray spectra, $-0.8$ and $-0.4$. \label{fig:chem}}
\end{centering}
\end{figure}

Figure \ref{fig:chem} shows the chemistry model results in terms of HCO$^+$ and N$_2$H$^+$ radial column density profiles for the full grid of models i.e. the grid of $4\times5$ X-ray and cosmic ray flux levels. Increasing X-ray hardness generally increases the HCO$^+$ and N$_2$H$^+$ column densities.  CR ionization does not strongly affect the HCO$^+$ column profile but does change the shape of N$_2$H$^+$, going from more centrally peaked column densities to radially flatter with increasing CR flux, similar to what was seen in the parameter study in \citet{cleeves2014par} for a generic T Tauri disk model.

\subsection{Constraints on Ionizing Agents}\label{sec:ionresults}
\begin{figure*}[ht]
\begin{centering}
\includegraphics[width=0.8\textwidth]{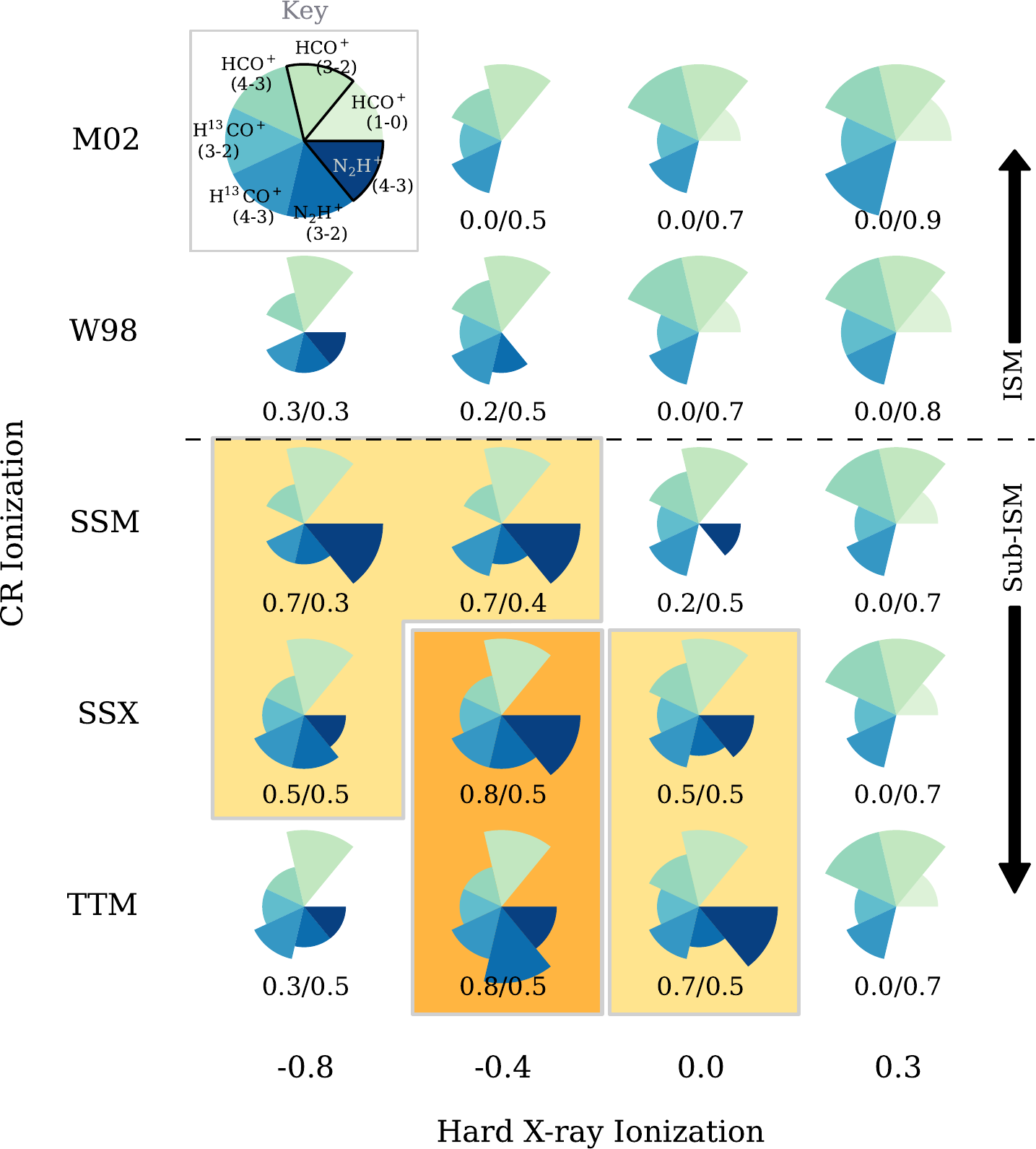} 
\caption{Ionization model goodness-of-fit for the lines indicated in the key (top-left).  The size of the wedge indicates how closely the model matches the data, where the largest wedge matches within $1\sigma$ (i.e., the best), the second and third smaller wedges indicate the model matches the data within 2$\sigma$ and 3$\sigma$ of the observations, respectively.  {\em No wedge implies the model and data do not match within $3\sigma$.}   The wedges that are boxed in black in the key indicate lines that are fit by their emission profile rather than their integrated line flux. Columns are X-ray ionization increasing from left to right (see Fig.~\ref{fig:X-rays}), while rows are CR ionization decreasing from the top down (see Table~\ref{tab:rates}).  The two numbers under each pie chart measure the goodness of fit to HCO$^+$ and N$_2$H$^+$ line fluxes, respectively. We box in orange the best fit models, where the darkest orange corresponds to the best-fit model. \label{fig:pie}}
\end{centering}
\end{figure*}

From the chemical models discussed in the previous section and the emission line analysis detailed in \S\ref{sec:modresults}, we can now compare the observations in Table~\ref{tab:data} to the simulated molecular emission.  The agreement between models and observed data is evaluated based on disk integrated fluxes for all lines except HCO$^+$ ($3-2$) and N$_2$H$^+$ ($4-3$), where the latter are fit based upon their radial emission profile (\S\ref{sec:lime}). The results of this comparison are presented in Figure~\ref{fig:pie}.  For each considered emission line, agreement is classified as within 1$\sigma$, 2$\sigma$, 3$\sigma$ or as a poor fit between the observation and model. We also classify the overall quality of fit to each model in terms of 1) midplane ionization as probed by N$_2$H$^+$ lines and 2) surface ionization tracers, i.e. HCO$^+$ lines. Finally, the total quality of fit is determined based on the fit to midplane and surface ionization levels weighted equally, which implies that individual N$_2$H$^+$ lines are weighted more heavily than individual HCO$^+$ lines for the final metric.  Based upon these numbers, the best fit and acceptable fit models to both surface and midplane tracers are boxed in Figure~\ref{fig:pie}.

Interestingly, all of the acceptable models have {\em sub-interstellar CR ionization rates}, $\zeta_{\rm CR}\lesssim10^{-18}$~s$^{-1}$.  The best fit models, SSX and TTM, have $\zeta_{\rm CR}\lesssim10^{-19}$~s$^{-1}$.  Additionally, these two models also have an X-ray spectral hardness ratio of $-0.4$, which is harder than TW Hya's quiescent X-ray state, $-0.8$, from the modelled observations of \citet{kastner1999}.  This result implies that the chemistry seems to reflect an elevated X-ray ionization state perhaps as a result of the well-characterized frequent flaring behavior of the star \citep{kastner2002,brickhouse2010}.   

\begin{figure*}[ht] 
\begin{centering}
\includegraphics[width=0.74\textwidth]{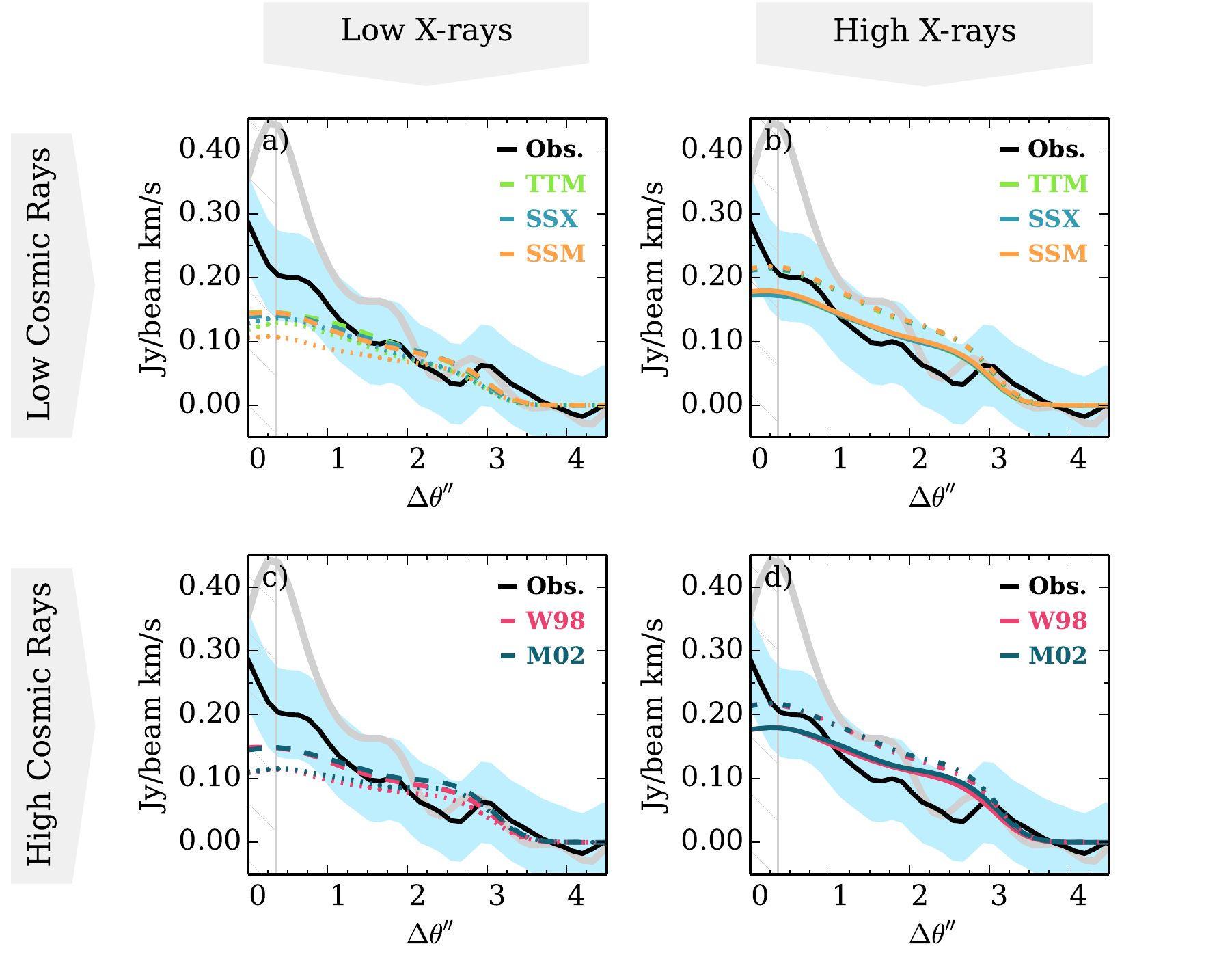}
\caption{Integrated line intensity profile of HCO$^+$ ($3-2$) versus angular distance from the star for the ionization models (colored lines) and the observed profile (solid black line).  We have divided the models into low, modulated CR ionization rate models (top) and high, interstellar CR ionization rate models (bottom). The line style indicates the X-ray spectral hardness ratio (Fig.~\ref{fig:X-rays}) where the two softer X-ray spectra are shown on the left panels: dotted ($-0.8$) and dashed ($-0.4$); and the two harder X-ray spectra are shown on the right:  solid (0.0) and dot-dashed (0.3).  The inner vertical region hatched in gray in the HCO$^+$ profiles designates the part of the disk that is contaminated by the asymmetric feature (shown as the solid gray profile), which is excluded from the fitting.  \label{fig:prof1} 
}
\end{centering}
\end{figure*}

\begin{figure*}[ht] 
\begin{centering}
\includegraphics[width=0.67\textwidth]{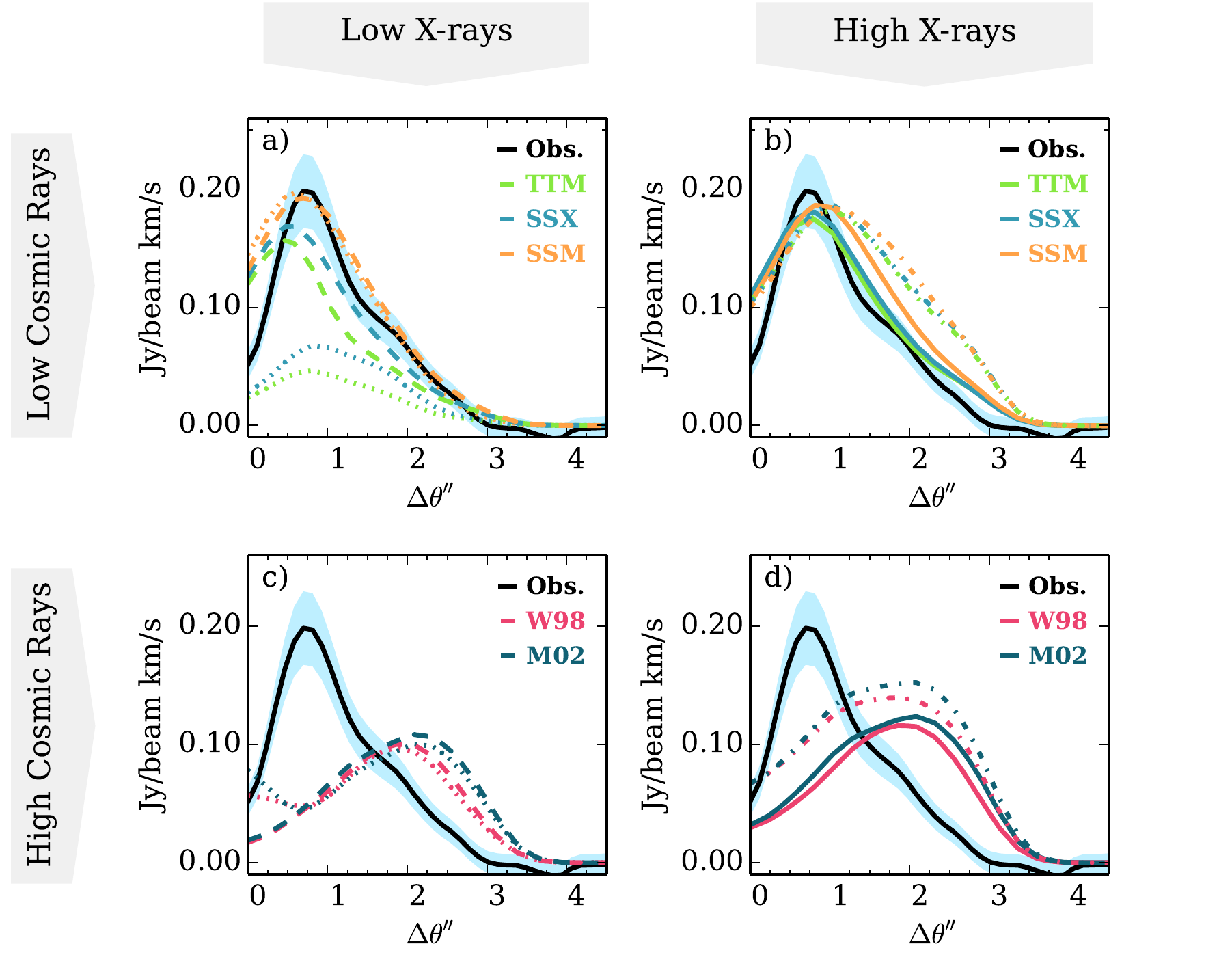} 
\caption{Integrated line intensity profile of N$_2$H$^+$ ($4-3$) versus angular distance from the star for the ionization models (colored lines) and the observed profile (solid black line).  The panels are divided up as in Figure~\ref{fig:prof1}.  The low CR models (top) are generally a better fit to the observed N$_2$H$^+$ ($4-3$) profile in tandem with a slightly harder X-ray profile, where the dashed line (top left panel, $-0.4$) or the solid line (top right panel, 0.0) provide a good fit. \label{fig:prof2}
}
\end{centering}
\end{figure*}

The quality of fit is especially apparent in the emission profile fitting of HCO$^+$ ($3-2$) and N$_2$H$^+$ ($4-3$), shown in Figures~\ref{fig:prof1} and \ref{fig:prof2} and described in \S\ref{sec:lime}.   To evaluate the model goodness of fit for the relevant wedges in Figure~\ref{fig:pie}, the HCO$^+$ ($3-2$) emission line models are convolved with a ($0.69''\times0.39''$) Gaussian beam, while the N$_2$H$^+$ ($4-3$) profile is measured from the CASA simulations discussed above.   Both the observations and models are averaged over deprojected annuli assuming an inclination of $i=7^\circ$.  The HCO$^+$ ($3-2$) profile  excludes the southwest quadrant from the annular averaging, along with the inner $0.3''$, due to the significant asymmetry present there. 

Most models reproduce the radial distribution of HCO$^+$ ($3-2$), while the N$_2$H$^+$ ($4-3$) spatial structure is more discriminatory between models.  Specifically, Figure~\ref{fig:prof2} shows that the N$_2$H$^+$ ($4-3$) ring discovered by \citet{qioberg2013} is  narrow and the emission drops off steeply beyond the peak at $\theta\sim0.7''$. These observed features are only reproduced by low, modulated CR models and moderately hardened X-ray spectra.  The broad N$_2$H$^+$ emission extending to large angular scales in the high CR case is due to non-thermal desorption of N$_2$ ice coupled with slower outer disk freeze-out times in the presence of CR ionization of H$_2$. This exclusion of high CR models is independent of the specifics of the assumed disk density and temperature profile (Appendix~\ref{app:addmod}).  There is thus very good agreement between the best-fit model derived from line flux comparisons and the spatial profiles of individual lines.

It is apparent from Figure~\ref{fig:pie} that the best fit models reproduce the N$_2$H$^+$ emission better than the HCO$^+$ emission on the whole.  The HCO$^+$ lines other than ($3-2$) are better fit by very hard X-ray spectra (HR: 0.3) or a high CR rate and HR of 0.0 or 0.3, and the models generally under predict the HCO$^+$ emission for all lines except the ($3-2$) transition. Some of this under prediction may be explained by the observed asymmetric excess in the ($3-2$) image (Fig.~\ref{fig:hcopdata}), which we were able to remove from the HCO$^+$ ($3-2$) flux before model-data comparison, but may contribute significantly to the other line fluxes. A second source of error is the uniform reduction of the CO abundance (see Appendix~\ref{sec:carbon}).  There could be substantial spatial structure in the gas-phase CO abundance distribution, which is the precursor for HCO$^+$ formation.  A third source of error is probably the details of the temperature structure, since the optically thin H$^{13}$CO$^+$ lines, which should trace the column well, are better reproduced by the model than the optically thick low-spatial resolution lines, which should trace the disk (surface) temperature. There is clearly a need for future detailed thermal structure modeling using more higher resolution HCO$^+$ and CO data that can be used to optimize the temperature and chemical/physical structure simultaneously.

\section{ Discussion}\label{sec:discussion}

\subsection{The Ionization Environment of TW Hya}
From the emission modeling, we find that the CR ionization rate is substantially suppressed at all disk radii, with especially strong limits on the outer disk.  One possible explanation is exclusion by young stellar winds as an analogue to the Solar System's heliosphere, i.e., a T-Tauriosphere.  As was shown in \citet{cleeves2013a} as well as \citet{svensmark2006,cohen2012}, winds from young stars should be efficient CR excluders if they operate over a large enough region of the disk.  Alternatively, if such winds are highly collimated by, e.g., magnetic fields, their covering fraction may be too small to shield the disk.  Whether disk winds are also capable of excluding CRs should be explored in future work. If stellar or disk winds are the  primary exclusion mechanism, they must operate well beyond the outer 200~AU gas disk radius, and the corresponding T-Tauriosphere, must be at least this large \citep[see discussion in][]{cleeves2013a}.  

CRs can also be repelled by mirroring via external magnetic fields linking the disk to the parent cloud, especially if there are irregularities in the field lines \citep{padovani2013,fatuzzo2014}.  Given TW Hya's relative isolation from molecular cloud material, it is uncertain whether such a large-scale environmental magnetic field exists.  Turbulent magnetic fields within the disk can also act as an additional source of energy loss for the CR particles if they are present \citep{dolginov1994}; however, the disk must be sufficiently turbulent (and thus MRI-active) to support such a configuration.  

The ionization rate derived for the midplane gas, $\zeta_{\rm CR}\lesssim10^{-19}$~s$^{-1}$, is, strictly speaking, a limit on {\em all} sources of ionization, including that due to SLRs and the scattered X-ray field in the midplane.  If TW Hya had similar SLR abundances to the Solar Nebula and without significant additional late-stage SLR pollution by massive stars \citep{adams2014}, the contribution from SLR ionization in TW Hya falls below the X-ray ionization rate in the midplane, or $\zeta_{\rm XR}=(2.3-10)\times10^{-20}$~s$^{-1}$, at $\gtrsim3$~Myr.  For our best fit model, the scattered X-rays alone exceed the contribution from SLR ionization; thus further constraining the ionization rate due to SLR ionization will be difficult.  Ionization tracers of the inner disk ($R<10$~AU) midplane may help put more stringent limits on the dense gas ionization in the region where X-rays are highly attenuated and SLRs may dominate.  Such small scales will be readily accessible by ALMA in the near future either with direct imaging or by using velocity information to spectrally resolve the inner disk.

 The overall picture of TW Hya's relatively high surface ionization rate and ion-poor midplane is  consistent with previous theoretical \citep[e.g.][]{gammie1996,glassgold1997,igea1999,semenov2004,semenov2010a,semenov2011} and observational \citep{kastner1997,qi2003,dutrey2007,oberg2011dm} studies of disk ionization.  With the IRAM Plateau de Bure Interferometry, \citet{dutrey2007} conducted a search for N$_2$H$^+$ and HCO$^+$ (and isotopologues) to measure the ionization state of three protoplanetary disks in a similar manner to the present paper.  While the PdBI data had lower resolution and lower signal-to-noise, the N$_2$H$^+$ observations of the three disks indicate a distinct drop at large radii, similar to what was found in the resolved TW Hya N$_2$H$^+$ $(4-3)$ observations (see Fig.~\ref{fig:prof2}).  Similar to our own findings \citep[c.f.][]{cleeves2014par}, the \citet{dutrey2007} chemical models with CRs show a rise in N$_2$H$^+$ column in the outer disk that is not present in the data (see their Figure~6).  We attribute this rise in the models to outer disk CR ionization and decreased ion-recombination efficiency.  While the lower signal-to-noise data presented in \citet{dutrey2007} cannot definitively point to a CR-poor environment, these observations may hint that TW Hya's low CR environment is a common feature of protoplanetary disks.

\subsection{Dead-zones, Dust Growth and Accretion}\label{sec:dust}
\begin{figure*}[ht]
\begin{centering}
\includegraphics[width=0.85\textwidth]{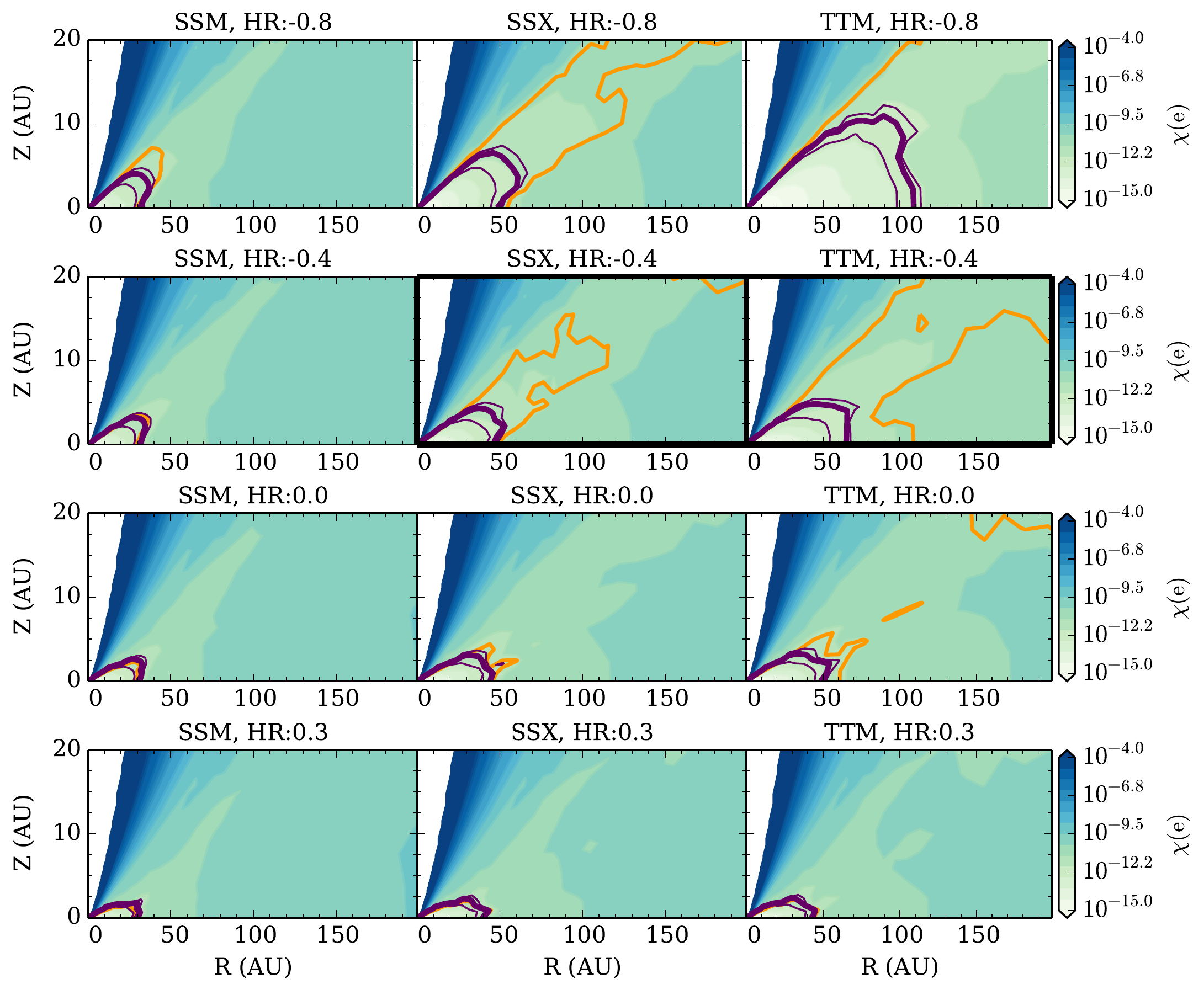}
\caption{Estimates of the dead-zone size as determined from the chemical calculations (the electron abundances at 1~Myr; see Section~\ref{sec:chem}).  The MRI-dead region is determined from the Reynolds number ($Re < 3000$, dark purple) and from the ambipolar diffusion parameter ($Am < 0.1$, orange) following \citet{pbc2011a} and \citet{cleeves2013a}.  The ambipolar diffusion criteria formally depends upon the strength of the magnetic field, which is unconstrained here, and so we emphasize that the Reynolds number (purple) provides the minimum predicted dead-zone size in this formalism.  For the best fit models, the dead-zone extends to $R=50$~AU and $R=65$~AU for the SSX and TTM models (black-boxed panels). \label{fig:deadzone}}
\end{centering}
\end{figure*}

From our best fit ionization models we can estimate the size of the magneto-rotational instability (MRI) ``dead'' regions in the disk in a similar fashion to \citet{cleeves2013a}.  In contrast to \citet{cleeves2013a}, we directly measure the ionization fraction from the chemical models, where the electron abundance is exactly equal to the ion abundance for a charge neutral disk.  From the spatial ion abundances (equivalently the ion fraction), we estimate the magnetic Reynolds number, $Re$, and ambipolar diffusion coefficient, $Am$, to determine the approximate location of the MRI active versus dead layers.  In Figure~\ref{fig:deadzone}, we plot the electron abundance as filled contours.  

We note that the electron abundance is slightly higher at the outer disk midplane in the present models than the steady-state estimate, $\chi_e=\sqrt{\zeta \alpha n({\rm H_2})}$, predicts because of the low densities and correspondingly longer recombination time-scales in the outer disk.  We choose a critical $Re$ for the disk to be dead as $Re\le3300$ (where MRI active lies outside of the area inclosed by the thick purple line), but we also show $Re=1000$ and $Re=5000$ on the same plot, demonstrating that the $Re$-determined dead-zone is not very sensitive to this choice.  Critical $Am<0.1$   (below which the disk is stable against the MRI)  is indicated by the region inside of the orange lines, and is mainly confined to a layer between the highly ionized surface and the outer disk midplane outside of $R>100$~AU (see grey contours of Fig.~\ref{fig:deadzone}).  Formally the disk must satisfy both $Am>0.1$ and $Re>3300$ to be active; however, the importance of the $Am$ criteria is affected by the strength and direction of the disk magnetic field, which is unconstrained here.  Therefore we focus on $Re$ to approximately estimate the minimum region where the disk is dead versus active.

For the two best fit models, SSX and TTM ($\rm HR= -0.4$), the $Re$ criteria results in a MRI dead region that extends from the inner disk out to 50~AU and 65~AU in the midplane (see Figure~\ref{fig:deadzone}).  This specific zone is of particular significance as it coincides with a region of concentrated large (mm/cm-sized) grains.  \citet{wilner2000} reported observations of TW Hya with the Very Large Array (VLA) where 7~mm continuum emission was found to be concentrated within a $R\sim50$~AU radius along with an unresolved 3.6~cm detection within a $\sim1.12''\times0.93''$ ($62\times51$~AU) beam. \citet{wilner2005} reported 3.6~cm continuum observations at higher spatial ($0.15''$) resolution and found that the emission was concentrated in a region of tens of AU in size.  Deep 870~$\mu$m observations with the Submillimeter Array (SMA) show similar morphology, where grains are concentrated within $R\sim60$~AU of the parent star \citep{andrews2012}, also seen in CARMA 1.3~mm observations where \citet{isella2009} find a disk radius of $\sim73$~AU.

The co-spatial location of the dead-zone with the large grains is perhaps not coincidental. \citet{gressel2012} argue that the decrease in turbulent activity within a dead-zone leads to the survival and growth embedded planetesimals, and thus may facilitate the growth of planets.  Additionally \citet{matsumura2005} and \citet{matsumura2007} argue that the presence of dead-zones may also halt rapid inward migration, leading to long-term survival of the planetesimal bodies.  We note there are certainly other explanations for the dust concentration and the sharpness of the mm-grain outer edge, in particular as a natural outcome of the velocity dependence of radial drift \citep[c.f.,][]{birnstiel2014}; however, a change in the disk turbulent properties resulting from a dead-zone may facilitate further dust growth via sticking collisions within this region.

If such  substantial dead-zone exists, it may also pose a problem for disk accretion onto the central star.  TW Hya is a relatively low accretor \citep[$\sim10^{-9}$~M$_\odot$~yr$^{-1}$;][]{alencar2002,herczeg2004,ingleby2013} supporting the continued existence of its gas-disk at a relatively old age; however, accretion must nonetheless proceed.  \citet{gammie1996} found that accretion can be sustained within a layer closer to the disk surface ($\Sigma_g<100$~g~cm$^{-2}$, limited by CR penetration) and in this case, estimate an accretion rate of $\dot{M}\sim10^{-8}$~M$_\odot$~yr$^{-1}$.  Parts of the dead-zone can also potentially become ``undead'' when reactivated by adjacent turbulent gas \citep{turner2008}, in which case the accretion flow in these regions is reduced compared to the active zone but is not zero.   \citet{lesur2014} found inclusion of the Hall effect in MHD shearing box simulations can support efficient accretion onto the star $\dot{M}\lesssim10^{-7}$~M$_\odot$~yr$^{-1}$, even for poorly ionized disks, with laminar flow through the dead zone, supporting the hypothesis that dead-zones may be beneficial to dust growth and planetesimal formation.  Thus the dead-zone may not entirely inhibit disk accretion, though the relationship between massive, extended dead-zones and mass/angular momentum transfer should be studied further.  

\subsection{HCO$^+$ Asymmetry}\label{sec:asymmetry}

Our new SMA observations reveals a significant small-scale asymmetric emission excess of HCO$^+$. While a detailed analysis of this particular feature is left to future work, there are a few plausible explanations for its origin.  The enhancement may be related to a local change in vertical structure, either due to a asymmetric pressure bump or the crest of a spiral arm.  The local increase in scale-height increases the surface area over which the disk can intercept stellar irradiation \citep{jangcondell2008,jangcondell2009} including that of X-rays, leading to a local over-production of HCO$^+$.  This scenario would also explain why the same feature does not appear in N$_2$H$^+$, which originates deeper in the disk, hidden from surface irregularities.  Alternatively, the presence of an accreting protoplanet still embedded within the disk might locally heat the gas, increasing the local CO abundance near the midplane and produce deep HCO$^+$; however, the ($3-2$) transition is likely optically thick and does not directly trace the dense gas where planets are expected to form.  The feature may also be temporal in nature, where the same stellar X-ray flares that are known to occur with some frequency may also be related to energetic expulsions from the central star, i.e., coronal mass ejections, that may impinge upon and ionize disk gas directly.  All of these scenarios are worth further explanation are beyond the scope of this paper, but hint at interesting chemical structure in the TW Hya disk that should be followed up with high resolution observations.

\section{Conclusions}\label{sec:conclusions} 

We have constrained the ionization environment of the TW Hya disk using molecular ion observations and a calibrated physical model of the TW Hya dust and gas circumstellar disk, on which we vary the incident ionizing flux of X-rays and CRs.  We find that the ionization rate due to CRs is quite low ($\zeta_{\rm H_2}\lesssim10^{-19}$~s$^{-1}$), and that the X-ray flares seem to have a lasting chemical effect on the disk.
We note that the limit for the CR rate is more than two orders of magnitude less than that derived for the dense interstellar medium.  We emphasize that the particular mechanism via which CR exclusion happens does not matter, but that the chemistry indicates that the CR flux is significantly lower than the canonical values throughout the full radial extent of the disk.  The main results of this paper are summarized as follows:
\begin{enumerate}
\item The total outer disk ionization rate in the disk midplane is below $\zeta_{\rm H_2}\lesssim10^{-19}$~s$^{-1}$.  This values formally puts limits on the combined CR, SLR and X-ray ionization rate throughout the disk midplane.  X-rays at the $\zeta_{\rm XR}\sim10^{-19}$~s$^{-1}$ level likely dominate the inner disk midplane and perhaps the outer disk at $2.3\times10^{-20}$~s$^{-1}$.  Due to the likely dominant contribution from scattered stellar X-rays at the midplane, it will be difficult to measure the CR and SLR ionization rate in the TW Hya disk directly.  This limit implies that the CR ionization rate in the outer disk is at least two orders of magnitude below that of the ISM.
\item The HCO$^+$ traces a slightly flared state of TW Hya (HR: $-0.4$) rather than the quiescent X-ray spectrum (HR: $-0.8$).  Additional detailed modeling of the thermal structure with resolved CO observations will help improve the fit to the optically thick HCO$^+$ lines, including the ($1-0$) rotational transitions.
\item  We make the first observational prediction of the dead-zone size, and based on the magnetic Reynolds' number, the expected dead-zone coincides with a region of known large grain concentration in the disk out to 60~AU.  A dead-zone of this size is consistent with the long lifetime of the gas disk in this system.
\item The HCO$^+$ ($3-2$) emission reveals slight asymmetry, which alone contributes $\sim20\%$ of the overall flux.  Resolved observations, where available, are thus extremely helpful when trying to understand the bulk disk properties. 
\end{enumerate}

In closing, we note that this study provides the strongest constraints to date on the sources of ionization in protoplanetary disks (at least for those constraints derived from imaging observations of ionized molecular species).  A wide variety of ionization sources are thought to contribute, including stellar X-rays, SLRs, CRs, and ionizing radiation from the background star-forming environment. This work shows observationally that CRs can indeed be excluded from disks, as proposed previously on theoretical grounds \citep[see, e.g.,][]{cleeves2013a,padovani2013,fatuzzo2014}, and provides an estimate for the extent of the dead zone (compare \citet{gammie1996} with \citet{igea1999}).  Given the
availability of new instruments and facilities, this study marks only the beginning of an important research program that will provide future observational constraints on disk physics.

\acknowledgements{The authors wish to thank the anonymous referee and editor for their helpful comments. LIC and EAB acknowledge grant AST-1008800 and KO acknowledges support from the Simons Collaboration on the Origins of Life (SCOL) program. CQ acknowledges grant NASA Origins of Solar Systems, Grant Number NNX11AK63G.  The Atacama Large Millimeter/submillimeter Array (ALMA), an international astronomy facility, is a partnership of Europe, North America and East Asia in cooperation with the Republic of Chile. This paper makes use of the following ALMA Science Verification data: ADS/JAO.ALMA\#2011.0.00001.SV.}


\appendix
\section{Physical Structure}\label{app:mod}
\subsection{Dust Model}\label{sec:dustmod}
We calibrate the disk physical density and temperature structure by fitting TW Hya's spectral energy distribution (SED).  References for the  SED photometry are provided in the Figure~\ref{fig:SED} caption, originally compiled by \citet{andrews2012}.  We adopt the same parametric density profile presented in \citet{cleeves2013a} Eqs.~(1-4), adapted from \citet{andrews2011}.  In essence, the gas and dust surface densities, $\Sigma_{g,d}$, are described by radial power-laws with an exponential taper, while the density, $\rho_{g,d}$, is taken to be  vertically Gaussian.  Moreover, we break the dust into two populations, one of small ``atmosphere'' grains with radii $r_g=0.005-1\mu$m distributed over the full (gas) scale height of the disk, and a second of larger midplane grains, $r_g=0.005-1$mm, concentrated near the midplane. The former contains 10\% of the total dust mass, while the latter contains the remaining 90\%.  This larger population is designed to simulate the effects of settling due to grain-growth, a feature common of observed protoplanetary disks \citep{furlan2006}. 
\begin{figure}[ht!]
\begin{centering}
\includegraphics[width=0.47\textwidth]{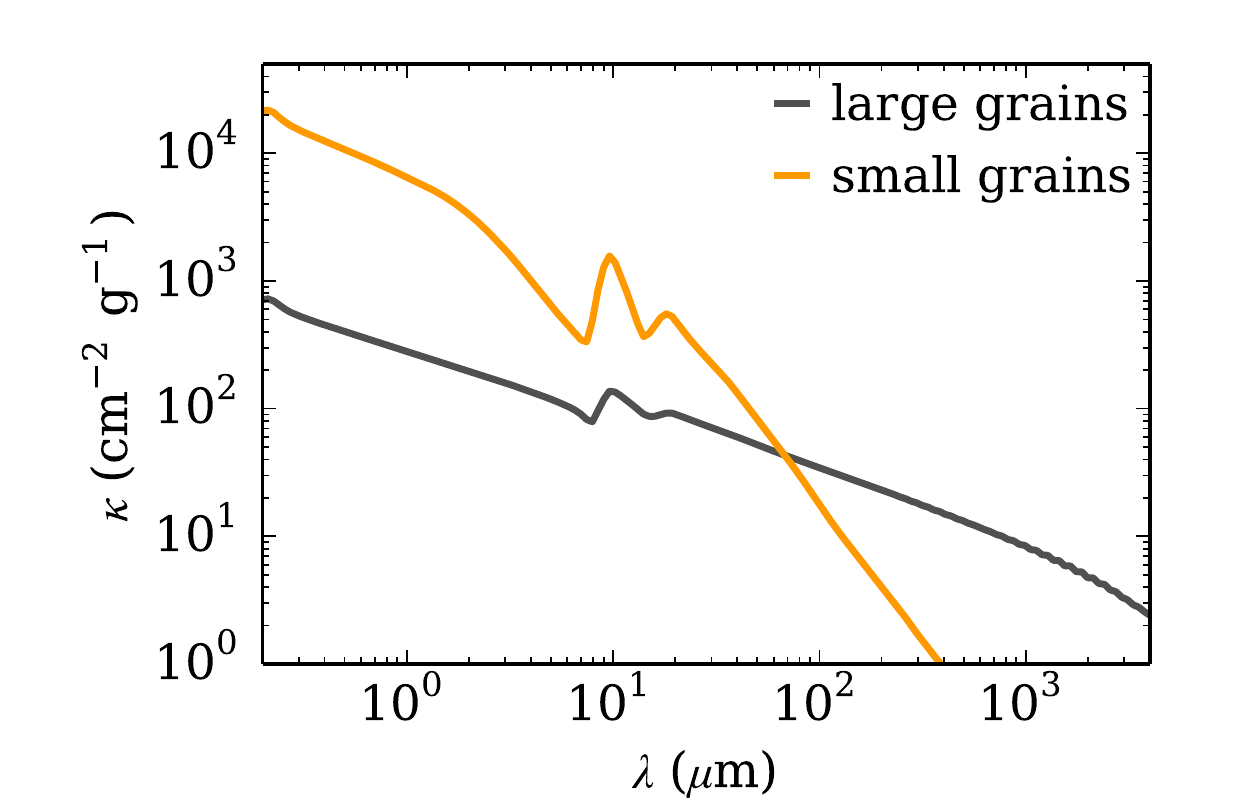}
\caption Dust opacities for the two dust populations used in our SED model. Small grains (atmosphere) have a maximum grain size of 1~$\mu$m, while the large grain population has a maximum wavelength of 1~mm. Opacities are plotted as a cross section per unit dust mass in grams. \label{fig:opa}
\end{centering}
\end{figure}
 Both large and small dust populations have an MRN size distribution \citep{mrn1977}, where the number of grains scales with the size of grain as $n_g\propto r_g^{-3.5}$.  We assume a mixed dust composition with 80\% astronomical silicates and 20\% graphite  by mass. The model opacities are shown in Figure~\ref{fig:opa}.
Similar to Eqs.~(1-4) of \citet{cleeves2013a}, our best fit model has a {\em total} dust surface density of
\begin{equation}
\Sigma_{d}(R)=0.04~{\rm g~cm^{-2}} \left( \frac{R}{R_c}  \right) ^{-1} \exp{ \left[ -\frac{R}{R_c} \right]},
\end{equation}
and a scale height for small grains (and gas) following
\begin{equation}
h(R)={\rm 15~AU}~{\Bigg(\frac{R}{R_c}\Bigg)}^{0.3}, \label{eq:hr}
\end{equation}
where the critical radius is $R_c=150$~AU.  The density of the small and large grain populations are described by
\begin{equation}
\rho_s (R,Z) = \frac{(1-f) \Sigma_d}{\sqrt{2 \pi}Rh} \exp{ \left[ - \frac{1}{2} \left(   \frac{Z}{h}  \right)^{2} \right]}, 
\end{equation} and
\begin{equation}
\rho_l (R,Z) = \frac{f\Sigma_d}{\sqrt{2 \pi}R  \chi h}  \exp{ \left[ - \frac{1}{2} \left(   \frac{Z}{\chi h}  \right)^{2} \right]},
\end{equation} 
respectively, where the fraction of mass in large grains is $f=0.9$ and the large grains concentrated over $\chi=0.2$ the scale height of the small grains, $h$.  $Z$ corresponds to the vertical distance from the midplane where $R$ and $Z$ are in cylindrical coordinates.

We use the radiative transfer code TORUS \citep{harries2000,harries2004,kurosawa2004,pinte2009} to calculate the dust temperatures and emergent SED assuming dust radiative equilibrium where heating is dominated by the central star.  We adopt the following stellar parameters for TW Hya: $T_{\rm eff}=4110$~K, $M_*=0.8$~M$_{\odot}$, $R_*=1.04$~R$_{\odot}$ \citep{andrews2012}.  The total mass of dust with grain sizes up to 1~mm in our best fit model  is $M_d=4\times10^{-4}$~M$_\odot$.  There may certainly be larger ``pebbles,'' boulders, or even planetesimals; however, the SED modeling is not sensitive to these.  We present the best fit SED model in Figure~\ref{fig:SED}. The corresponding dust density and temperature models are shown as solid contour lines in Figures~\ref{fig:model}a and \ref{fig:model}b.

\begin{figure}
\begin{centering}
\includegraphics[width=0.47\textwidth]{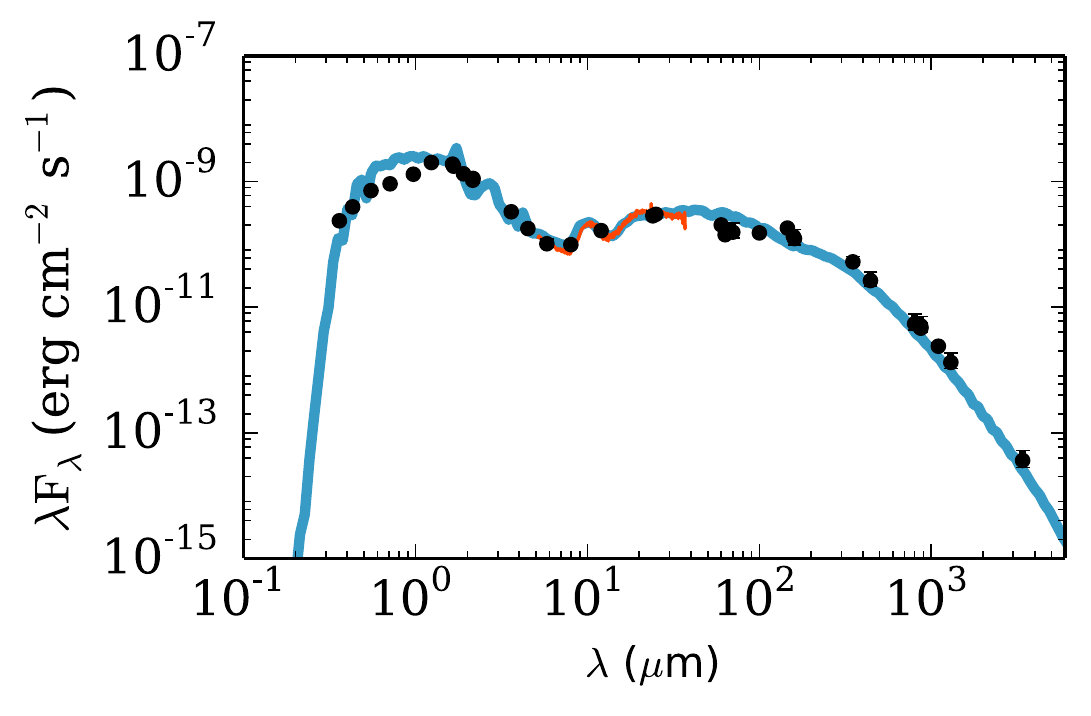}
\caption{Our best fit spectral energy distribution (blue line) from the combined star and disk of TW Hya.  Black points (with error bars) are individual photometric measurements taken from the literature \citep{weintraub2000,mekkaden1998,cutri2003,hartmann2005,low2005,thi2010a,andrews2012,weintraub1989,qi2004,wilner2003,wilner2000}. TW Hya's Spitzer IRS spectrum is over-plotted in orange \citep{uchida2004}.  
\label{fig:SED}}
\end{centering}
\end{figure}

\subsection{Gas Model}\label{sec:gas}
The SED fitting formally constrains the distribution of dust grains.  To constrain the total {\em gas} mass in the TW Hya protoplanetary disk, we use the {\em Herschel} detection of hydrogen deuteride, HD,  to directly probe the gas reservoir \citep{bergin2013}.  This is especially crucial for TW Hya's disk, as the more widely used molecular gas probe, CO, is measured to be depleted in warm gas, i.e., in addition to CO freeze-out \citep[][see also Sec.~\ref{sec:carbon}]{favre2013}.   Because HD does not freeze-out and the HD to H$_2$ ratio is approximately constant and well constrained within the local bubble \citep[HD/H$_2=3\pm0.2\times10^{-5}$;][]{linsky1998}, HD is a chemically ``simple'' species, differing mainly from H$_2$ with regards to UV self-shielding. 

To calibrate our model's gas mass using HD, we start from the dust-derived physical structure and assume the gas density and small grains are distributed over the same scale height and are vertically Gaussian (filled contours in Fig.~\ref{fig:model}a).  Furthermore, we must assume a gas-to-dust mass conversion factor where, because gas and dust are not uniformly vertically distributed due to settling, we compare  the vertically integrated quantities, $f_g=\Sigma_g/\Sigma_d$, with $f_g=100$ for dense interstellar gas.  

To calculate the HD emission, we must also estimate the  gas temperature and dust opacity at 112~$\micron$.  HD is particularly sensitive to the gas temperature  due to the high upper state energy of the $J=1$ rotational level.  In the upper layers of the disk, the gas temperature can exceed the dust temperature, i.e., become ``decoupled'' due to less efficient gas cooling. 
 We estimate the gas temperature by using a fitting function calibrated to thermochemical models of FUV heating from the central star (S. Bruderer, in private communication 2013). Based on a grid of physical structures and FUV field strengths, the detailed heating and cooling balance was solved \citep[e.g.,][]{bruderer2013} to determine the gas thermal decoupling from the dust in the disk atmosphere where $A_V=1-3$.  Based on these models, S. Bruderer estimates a parameterized fit to the gas temperature based on the local FUV strength and gas density.  We emphasize that the main results of this paper, the ions, are not sensitive to the gas temperatures as most of the molecular ion emission comes from deeper layers and, correspondingly, higher $A_V$.  To determine the FUV radiation field throughout the disk, we use a Monte Carlo technique to calculate the wavelength-dependent UV field including both absorption and scattering on dust, in addition to the transfer of Lyman-$\alpha$ photons \citep{bethell2011b}.  Special treatment of Lyman-$\alpha$ radiative transfer is important as these photons will resonantly scatter on atomic hydrogen atoms, greatly enhancing the scattered radiation field deep in the disk compared to the primarily forward-scattering dust.  Furthermore, Lyman-$\alpha$ carries $\sim85\%$ of the FUV flux below 2000~\AA\  \citep{herczeg2004}. The UV optical properties are taken from the dust model described in \S\ref{sec:dust}, and we use the measured TW~Hya FUV fluxes from \citet{herczeg2002,herczeg2004} assuming a distance of $d=55$~pc. The calculated FUV radiation field integrated over wavelength is shown in Figure~\ref{fig:model}d, and the resulting gas temperature structure is shown in Figure~\ref{fig:model}b as filled color contours.  For the most part $T_g=T_d$ (see contour lines), however the gas temperature becomes decoupled from the dust by more than $\Delta T>10$~K in the layer where $z/r\gtrsim0.4$.   

In addition to the gas temperature, the HD emission is sensitive to the vertical structure of the dust as the dust disk becomes optically thick at 112~$\mu$m. Consequently, the HD emission is sensitive to the specific dust opacity for which we must assume a single value.  The weighted average of the two large and small populations corresponds to an opacity per gram of dust of $\rm \kappa_{mix}(112~\mu m)=30$~cm$^{2}$~g$^{-1}$.  

From the gas and dust density and temperature model, we compute the baseline chemistry for HD  (see details of the chemical code in \S\ref{sec:chem}) to primarily determine how much HD is dissociated in the upper layers before self-shielding takes hold.  Because HD does not freeze-out, the HD abundance is effectively constant throughout the disk below the UV self shielding layer ($z/r\sim0.4$). From the calculated HD abundances, we then compute the emergent HD ($1-0$) line intensity assuming the emission is in LTE (see \S\ref{sec:lime} for details on the radiative transfer).  We then adjust the gas-to-dust ratio, $f_g$, until we find agreement with the observed HD flux, $\int{F_{\rm HD}dv} = 70.6\pm7.8$~Jy~km~s$^{-1}$ \citep{bergin2013}. 
With a vertically integrated gas-to-dust mass ratio of $f_g=\Sigma_g/\Sigma_d=75-100$ ($M_g=0.03-0.04$~M$_\odot$), we find good agreement with the observed value, where our 0.04~M$_\odot$ model predicts $\int{F_{\rm HD}dv}=76.6$~Jy~km~s$^{-1}$.  We note, however, that slightly less massive ($M_g=0.02$~M$_\odot$) but warmer disk or perhaps a more massive ($M_g=0.05$~M$_\odot$) but cooler disk can also reproduce the observations, so in the present framework, we find that TW Hya's gas mass is $M_g=0.04\pm0.02$~M$_\odot$. This value can be further refined with better observational constraints on the overall vertical density and thermal structure.  The gas mass derived here is slightly less than the mass provided by \citet{bergin2013}, $M_g>0.05$~M$_\odot$, and is chiefly due to differences in the gas temperature calculation and underlying disk model assumed.  

\section{Neutral Gas Constraints}\label{app:neutral}

\subsection{Neutral Gas Constraints: CO}\label{sec:carbon}
In the ISM, CO is the second-most abundant gas phase molecule after H$_2$.  In the ISM, CO has an abundance of $\rm \chi(CO)=10^{-4}$ and participates in a wide range of chemical reactions.  However, recent observations indicate that CO is substantially reduced in warm ($T\gtrsim20$~K) gas where the CO abundance relative to H$_2$ was found to be $\chi({\rm CO})=(1-10)\times10^{-6}$ \citep{andrews2012,favre2013}.  \citet{williams2014} indirectly confirm this finding by deriving a gas mass from CO isotopologue observations of $M_g = 5\times10^{-4}$, a factor of $\sim100$ less than the HD derived gas mass.  One potential explanation for this large CO deficit is through CO dissociation by He$^+$, where some fraction of the carbon from CO to be put into other neutral species with higher grain-surface binding energies than that of CO.  This process can happen at early stages prior to the formation of the disk, activated by CR ionization, or at later stages in the disk's warm molecular layer initiated by stellar X-ray ionization \citep{semenov2004,bergin2013}.  To robustly make predictions for HCO$^+$ abundances, which forms from CO via
\begin{equation}
\rm H_3^++CO\rightarrow HCO^++H_2,
\end{equation}
we must include the CO deficit in our model. We initially ran models where the initial (input) CO abundance is set at $\chi({\rm CO})=1\times10^{-6}$, and the rest of the carbon is put into strongly bound, carbon-bearing ices, e.g., methanol.  However, even in this instance the carbon in methanol ice was recycled back into gas phase CO in less than 1~Myr in the layers where UV photons are present.  Even when we artificially increased the binding energy of methanol, the carbon nevertheless made its way back into gas phase CO, and over-produced the observed CO emission (i.e., the CO abundance after 1~Myr was far too high to explain the observations).  

In the end, we found that the only way to reproduce the low CO abundance was to reduce the CO abundance and explicitly {\em not} put it into one of the existing network species, thereby net reducing the amount of {\em reactive} carbon.  The physical interpretation behind this finding is that the carbon no longer in gaseous CO has gone on to form something similar to macromolecular organic ices.  In the presence of UV irradiation, such material is less likely to non-thermally desorb, and are more likely to break up into radical ices, where the products remain on the grains (and are not returned to the gas) and are thought to be key to forming important biogenic organic material.  

\begin{figure}
\begin{centering}
\includegraphics[width=0.49\textwidth]{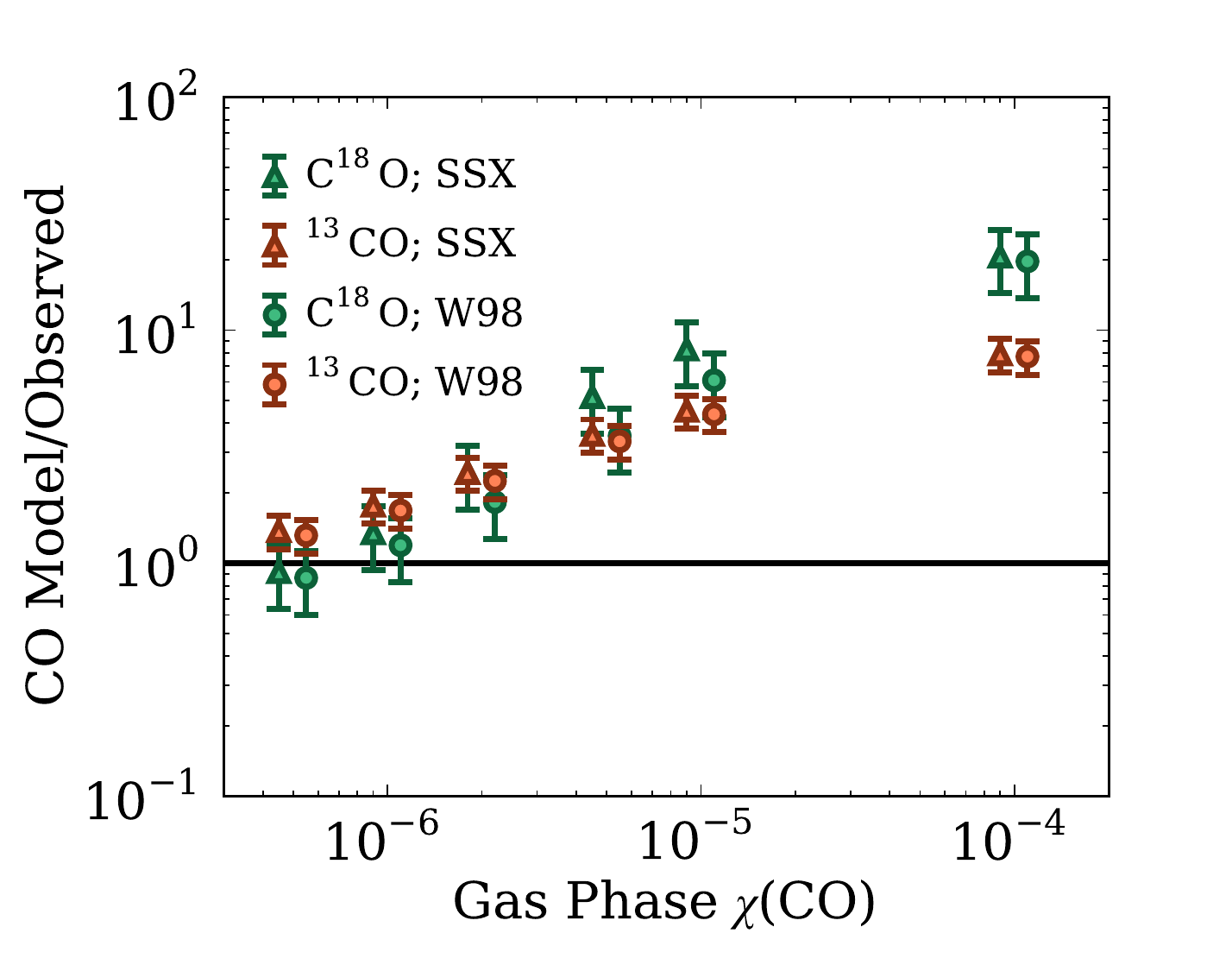}
\caption{Comparison between the simulated C$^{18}$O (2-1) and $^{13}$CO (2-1) line intensities and the observed values (see Table~\ref{tab:data}) for different gas-phase CO abundances. The canonical CO abundance, $\chi$(CO)~=~$10^{-4}$ over predicts the observed fluxes by approximately an order of magnitude and is also optically thick -- inconsistent with the observations.   A gas-phase CO abundance of $\chi$(CO)~=~$10^{-6}$ or perhaps lower is a better match to the data, in agreement with the findings of \citet{andrews2012} and \citet{favre2013}.   \label{fig:chico}}
\end{centering}
\end{figure}

Taking two models with different CR ionization rates (SSX and W98, \S\ref{sec:CRs}), we i) vary the {\em initial} CO abundance, ii) calculate the final CO abundance after 1~Myr, and iii) compute the $^{13}$CO and C$^{18}$O emergent line emission (see \S\ref{sec:lime}).  We compare these values to the observations \citep[][see also Table~\ref{tab:data}]{favre2013}.  Figure~\ref{fig:chico} shows the ratio of the observed flux to the model flux for different CO initial conditions.  The model which simultaneously best fits $^{13}$CO and C$^{18}$O is one where $\chi({\rm CO})=1\times10^{-6}$, though CO abundances between  $\chi({\rm CO})=(5-20)\times10^{-7}$ can fit either $^{13}$CO or C$^{18}$O.  The main differences between the present work and \citet{favre2013} is that we are using a new model and a different method to calculate gas temperature. We confirm the \citet{favre2013} result that the CO abundance is substantially lower than the canonical $\chi({\rm CO})=10^{-4}$ in the warm gas by approximately a factor of $\sim100$.

\subsection{Neutral Gas Constraints: HCN}\label{sec:nitro}
Motivated by the work of \citet[][submitted]{schwarz2014}, the initial abundances of nitrogen also play an important role in the chemical outcome of nitrogen bearing species, including HCN, NH$_3$ and N$_2$H$^+$.  Because N$_2$H$^+$ is simultaneously affected by ionization and CO abundance, the neutral species provide a cleaner test of the initial nitrogen abundances in this model.  Described in more detail in \citet[][submitted]{schwarz2014}, the final nitrogen molecular abundances are most sensitive to the following broad groupings of initial nitrogen reservoirs, namely the amount of NH$_3$ ice, N$_2$, and N and/or single-N containing molecules.  The species HCN has been previously detected in the TW Hya disk, and thus we use this molecule as a probe of the initial nitrogen portioning in the disk; however, we note that this is only one line and that future more detailed modeling and additional observations will greatly help but additional constraints on the nitrogen assay in disks.  

\begin{figure}
\begin{centering}
\includegraphics[width=0.41\textwidth]{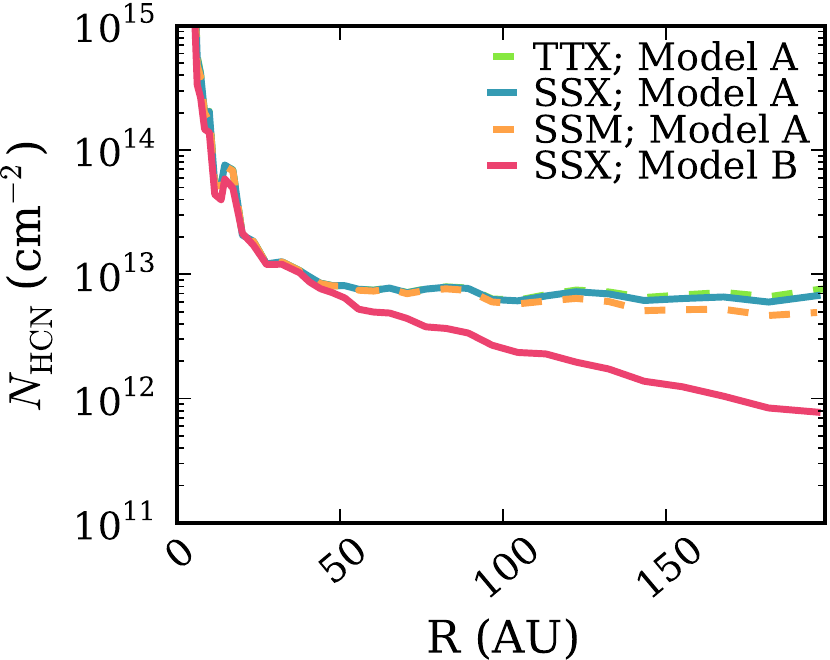}
\caption{The dependence on different initial nitrogen conditions of the HCN abundance after 1~Myr of chemical evolution.  The initial abundances relative to total H of major nitrogen-bearing species  (with abundance $\chi\ge10^{-7}$) are as follows.   Model A: $\chi$(NH$_3$ ice)~=~$9.16\times10^{-6}$, $\chi$(N$_2$)~=~$2.5\times10^{-6}$, and $\chi$(N)~=~$5.1\times10^{-6}$. Model B: $\chi$(NH$_3$ ice)~=~$7.91\times10^{-6}$, $\chi$(N$_2$)~=~$5.7\times10^{-6}$, and $\chi$(N)~=~$1\times10^{-7}$.  Variations between initial nitrogen abundances exceed variations due to different assumed ionization rates (dotted lines, Model A).  See Section~\ref{sec:nitro} for details. \label{fig:nit}}
\end{centering}
\end{figure}

In the midplane, since we use a static disk model, the chemistry is such that the abundance of NH$_3$ ice in the midplane is often very similar to that which was assumed initially, i.e., the ices are not rapidly reprocessed.  In this light, we can look to cometary NH$_3$/H$_2$O ratios to put an upper limit on the NH$_3$ ice abundance.  Typical ammonia abundances for comets have percentages of $0.1-1.5\%$ relative to water \citep{bockeleemorvan2004,biver2012}.  Evidence from protostellar NH$_3$ ice abundances, typically $\sim3\%$ \citep{oberg2011sp} and comets gives an approximate range for the amount of nitrogen locked up in ices and thus provides a crude handle on the nitrogen partitioning in the disk.   In Figure~\ref{fig:nit} we show two models considered where Model A has more ammonia ice (less reactive nitrogen)

The initial N$_2$ abundance directly affects N$_2$H$^+$; however, its effect is predictable.  The overall column density profile may shift up or down, but the shape of the N$_2$H$^+$ column density versus radius stays the same.  The N$_2$ and NH$_3$ binding energies assumed in the model are 1220~K and 3080~K, respectively. The final nitrogen abundances determined from the HCN emission are listed in Table~\ref{tab:abun}. 

\begin{deluxetable}{llll} 
\tablecolumns{4} 
\tablewidth{0pt}
\tablecaption{Chemical model initial abundances relative to total number of H-atoms. \label{tab:abun}}
\tablehead{  
Species &  $\chi$ & Species & $\chi$}
\startdata
H$_2$ & $5.00\times10^{-1}$ & H$_2$O(gr) & $2.50\times10^{-4}$ \\
HDO(gr) & $1.00\times10^{-8}$ & He & $1.40\times10^{-1}$ \\
CN & $6.60\times10^{-8}$ & HCN\footnote{See \S\ref{sec:nitro}.}   & $1.00\times10^{-8}$\\
N & $5.10\times10^{-6}$ & NH$_3$(gr) & $9.90\times10^{-6}$ \\
N$_2$\footnote{See \S\ref{sec:nitro}.}  & $1.00\times10^{-6}$ & H$_3^+$ & $1.00\times10^{-8}$ \\
CS & $4.00\times10^{-9}$ & SO & $5.00\times10^{-9}$ \\
Si$^+$ & $1.00\times10^{-11}$ & S$^+$ & $1.00\times10^{-11}$ \\
Mg$^+$ & $1.00\times10^{-11}$ & Fe$^+$ & $1.00\times10^{-11}$ \\
C$^+$ & $1.00\times10^{-9}$ & CH$_4$ & $1.00\times10^{-7}$ \\
Grain & $6.00\times10^{-12}$ & CO\footnote{See \S\ref{sec:carbon}.}  & $1.00\times10^{-6}$ \\
C & $7.00\times10^{-7}$ & HCO$^+$ & $9.00\times10^{-9}$ \\
HD & $1.50\times10^{-5}$ & H$_2$D$^+$ & $1.30\times10^{-10}$ \\
HD$_2^+$ & $1.00\times10^{-10}$ & D$_3^+$ & $2.00\times10^{-10}$ \\
C$_2$H & $8.00\times10^{-9}$ &  & 
\enddata
\end{deluxetable}

\section{Model Comparison}\label{app:addmod}
It is important to quantify the model dependency of these results.  To determine how our results depend on the disk physical structure, we repeat our experiment using the detailed model of \citet{andrews2012}.  The \citet{andrews2012} model fits the dust distribution in detail, and fits the CO $(3-2)$ profile.  Because CO ($3-2$) is thick, the gas model primarily reflects the disk temperature profile versus radius.  However, the best fit model in that work found a significantly smaller taper radius for the gas disk, i.e., the critical radius, where $r_c=35$~AU for the sA model, which could still fit the distributed CO gas out to 200~AU.  The model in the present work, for comparison, has a taper at $r_c = 150$~AU, dropping off instead at the edge of the CO and scattered light disk.  

\begin{figure*}[ht]
\begin{centering}
\includegraphics[width=0.76\textwidth]{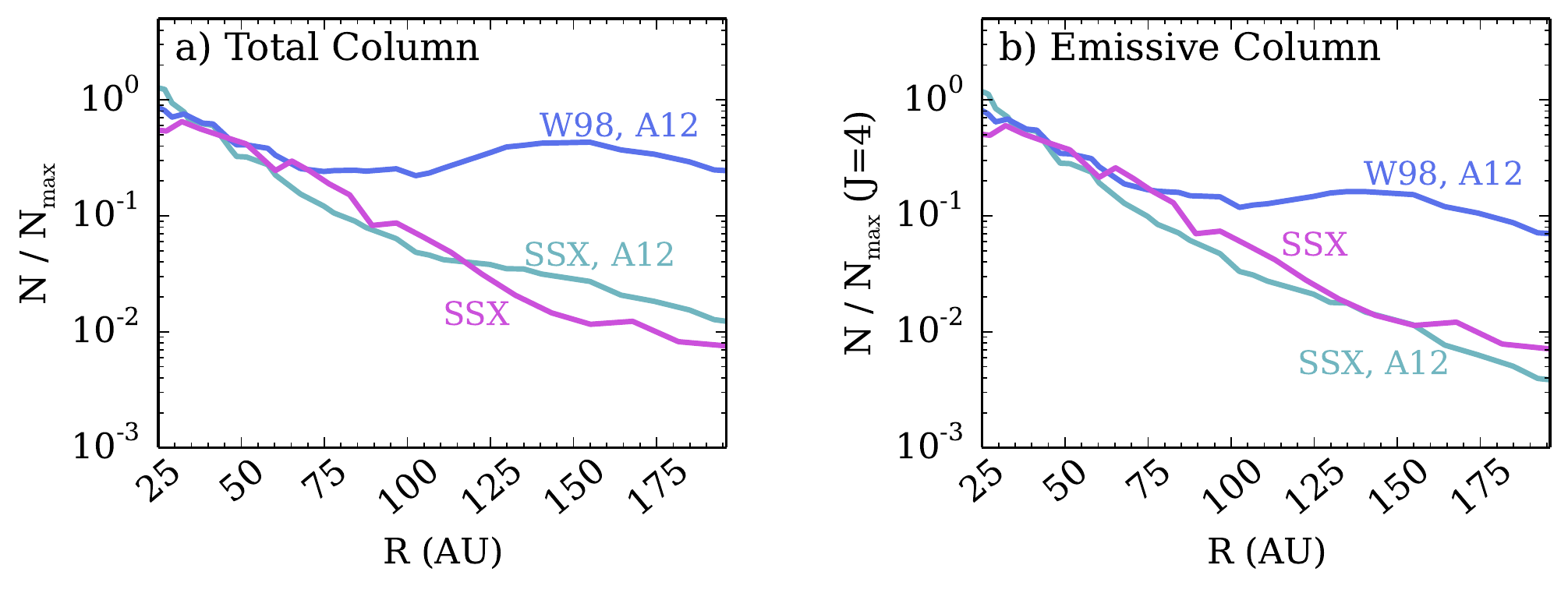}
\caption{Chemical model comparison between the disk model assumed in this work (labeled just SSX) and the results for a different underlying disk model from \citet{andrews2012} (lines labeled A12).  We find good agreement in the overall slope of the column densities for the low CR ionization model and can exclude the high CR ionization model, W98, which over predicts the outer disk by over an order of magnitude.  \label{fig:sacomp}}
\end{centering} 
\end{figure*}

Consequently, there is a substantial difference in the mass distribution between the two models, and the disk mass itself (which is an order of magnitude smaller for the \citet{andrews2012} model). The mass/density difference is most pronounced at the outer disk, at the same radii where N$_2$H$^+$ ($4-3$) is dropping off.  By comparing the two models, we can test whether or not the N$_2$H$^+$ distribution is a mass effect or an ionization effect.  In Figure~\ref{fig:sacomp}, we show the normalized N$_2$H$^+$ column density for the outer disk (Fig.~\ref{fig:sacomp}a) and the column density of the population of N$_2$H$^+$ in the $J=4$ upper state (which is more closely related to the line emissivity).  We have normalized the columns because we have not done any additional chemical calibration or mass calibration for the \citet{andrews2011} model as were done in the main paper, and so there is an overall offset between the two models.  From these tests we find that the overall slope of the N$_2$H$^+$ column density and emissive column density agree well with the results of the main paper for the SSX (reduced CR ionization model), and that even with the reduced outer disk mass in the \citet{andrews2011} model, the N$_2$H$^+$ emission distribution is too flat to explain the observations.  This behavior is a natural consequence of the drop in outer disk density, where the loss of mass acts to {\em reduce} the recombination efficiency of ions, and thus there is higher fractional abundance of ions, including N$_2$H$^+$, than in our model, which has higher outer disk recombination due to the higher outer disk mass in the present paper.  Thus the N$_2$H$^+$ profile cannot be attributed to a mass effect, and that a reduced CR ionization rate does a better job of explaining the emission distribution for both physical structures.

\end{document}